\documentclass{article}

\usepackage{arxiv}

\usepackage[utf8]{inputenc} 
\usepackage[T1]{fontenc}    
\usepackage{hyperref}       
\usepackage{url}            
\usepackage{booktabs}       
\usepackage{amsfonts}       
\usepackage{nicefrac}       
\usepackage{microtype}      
\usepackage{lipsum}
\usepackage{graphicx}
\usepackage{amsmath}
\usepackage{threeparttable}
\usepackage{placeins} 
\usepackage{subcaption}
\usepackage{float}

\title{Towards 3D Semantic Image Synthesis for Medical Imaging}

\author{
  Wenwu Tang\\
  Institute of Signal Processing and System Theory\\
  University of Stuttgart\\
  Stuttgart, 70569 \\
  \texttt{wenwutang2000@gmail.com} \\
   \And
 Khaled Seyam \\
  Institute of Signal Processing and System Theory\\
  University of Stuttgart\\
  Stuttgart, 70569 \\
  \texttt{khaled.seyam@iss.uni-stuttgart.de} \\
      \And
 Bin Yang \\
  Institute of Signal Processing and System Theory\\
  University of Stuttgart\\
  Stuttgart, 70569 \\
  \texttt{bin.yang@iss.uni-stuttgart.de} \\
}

\begin{document}
\maketitle

\begin{abstract}
In the medical domain, acquiring large datasets is challenging due to both accessibility issues and stringent privacy regulations. Consequently, data availability
and privacy protection are major obstacles to applying machine learning in medical imaging. To address this, our study proposes the Med-LSDM (Latent Semantic Diffusion Model), which operates directly in the 3D domain and leverages de-identified semantic maps to generate synthetic data as a method of privacy preservation and data augmentation. Unlike many existing methods that focus on generating 2D slices, Med-LSDM is designed specifically for 3D semantic image synthesis, making it well-suited for applications requiring full volumetric data. Med-LSDM incorporates a guiding mechanism that controls the 3D image generation process by applying a diffusion model within the latent space of a pre-trained VQ-GAN. By operating in the compressed latent space, the model significantly reduces computational complexity while still preserving critical 3D spatial details. Our approach demonstrates strong performance in 3D semantic medical image synthesis, achieving a 3D-FID score of 0.0054 on the conditional Duke Breast dataset and similar Dice scores (0.70964) to those of real images (0.71496). These results demonstrate that the synthetic data from our model have a small domain gap with real data and are useful for data augmentation.

\end{abstract}

\section{Introduction}

Despite the remarkable progress of deep learning (DL) in medical imaging, including tasks such as image analysis, diagnosis, and treatment planning \cite{MedicalImageAnalysis, Litjens_2017, he2022transformersmedicalimageanalysis}, the field continues to face fundamental challenges. One of the most critical issues is the scarcity of large, diverse, and high-quality labeled datasets. Medical data collection and annotation are inherently constrained by factors such as the high cost of expert labeling, data heterogeneity across imaging modalities, and strict privacy regulations governing patient information \cite{moeskops2022deep, luca2022impact, prevedello2019challenges}. These limitations hinder the ability of conventional deep learning models to generalize effectively, ultimately restricting their clinical applicability.

\begin{figure}[ht]
	\centering
	\includegraphics[width=0.8\linewidth]{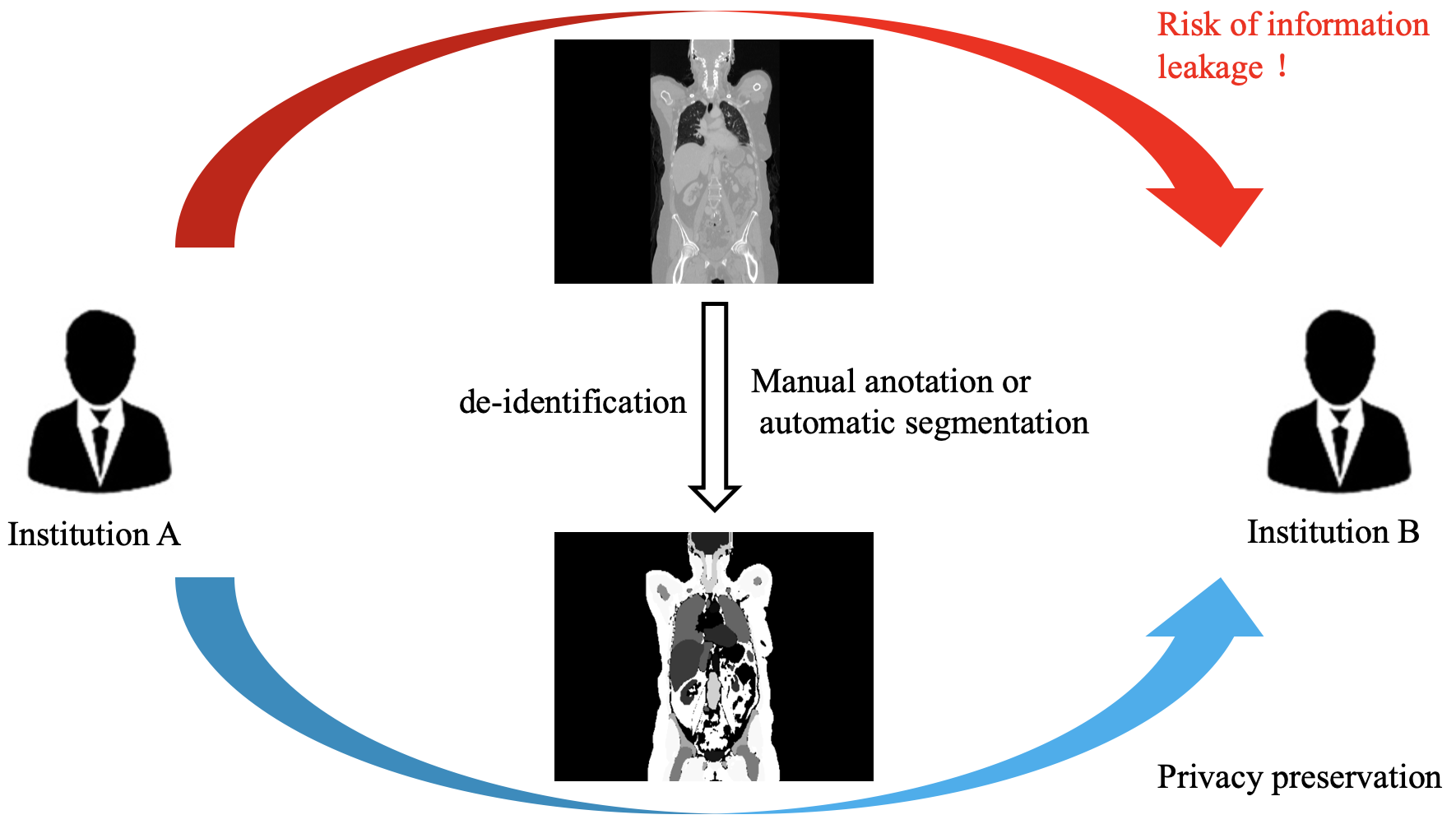} 
	\caption{Comparison of anonymous and non-anonymous data sharing}
	\label{fig:privacy}
\end{figure}

Generative models have emerged as a promising solution to mitigate these challenges. In particular, Generative Adversarial Networks (GANs) and Denoising Diffusion Probabilistic Models (DDPMs) \cite{ho2020denoisingdiffusionprobabilisticmodels} have shown great potential in augmenting medical datasets by synthesizing realistic, high-fidelity medical images \cite{zhang2021shifting, chen2021gan_review}. By enriching training datasets with synthetic samples, generative models enhance model generalization and robustness, particularly in data-limited scenarios. Furthermore, these models contribute to privacy preservation by enabling the generation of de-identified synthetic data, which mitigates the risk of patient re-identification while maintaining the clinical relevance of the data. Techniques such as label-to-image synthesis leverage high-level semantic information (e.g., segmentation maps or anatomical labels) to generate realistic medical images without directly exposing sensitive patient data. This approach aligns with stringent privacy frameworks, such as HIPAA \cite{hippa}, and fosters secure data sharing and collaboration between medical institutions \cite{Ma_2024}. As shown in Figure \ref{fig:privacy}, sharing semantic labels or synthetic images instead of raw patient data allows cross-institutional research while minimizing privacy risks.

\begin{figure}[h]
    \centering
         \begin{minipage}{0.15\textwidth}
        \centering
        \includegraphics[width=\textwidth]{ 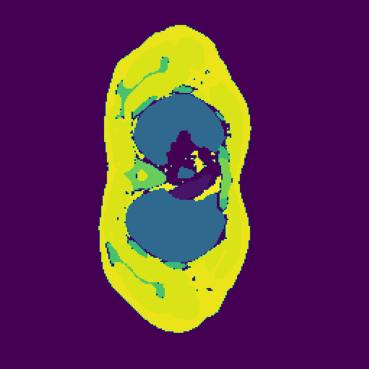}
    \end{minipage}%
    \begin{minipage}{0.15\textwidth}
        \centering
        \includegraphics[width=\textwidth]{ 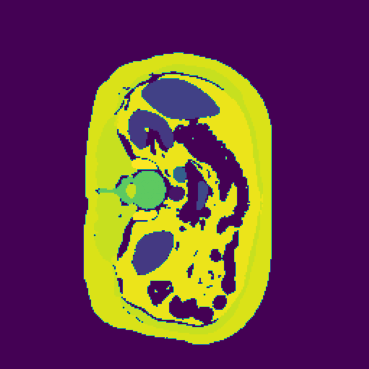} 
    \end{minipage}%
    \begin{minipage}{0.15\textwidth}
        \centering
        \includegraphics[width=\textwidth]{ 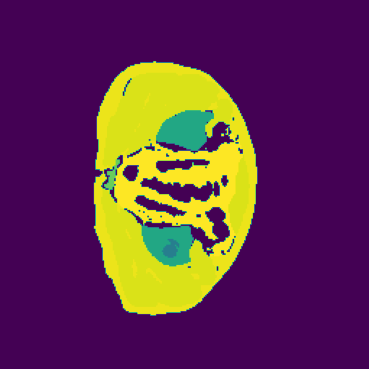}
    \end{minipage}%
      \begin{minipage}{0.15\textwidth}
        \centering
        \includegraphics[width=\textwidth]{ 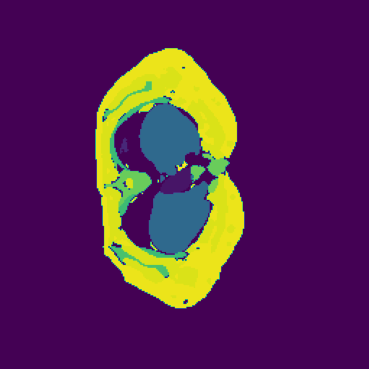}
    \end{minipage}%
      \begin{minipage}{0.15\textwidth}
        \centering
         \includegraphics[width=\textwidth]{ 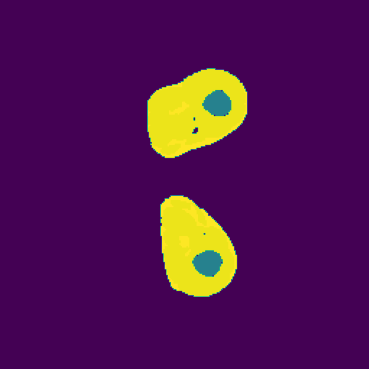}
    \end{minipage}%

    \begin{minipage}{0.15\textwidth}
        \centering
        \includegraphics[width=\textwidth]{ 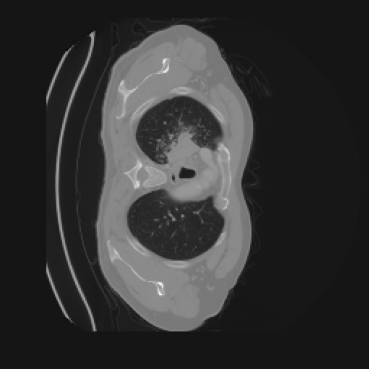}
    \end{minipage}%
    \begin{minipage}{0.15\textwidth}
        \centering
        \includegraphics[width=\textwidth]{ 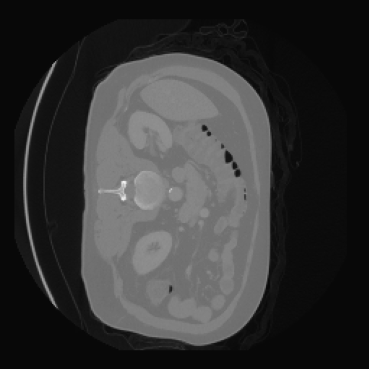} 
    \end{minipage}%
    \begin{minipage}{0.15\textwidth}
        \centering
        \includegraphics[width=\textwidth]{ 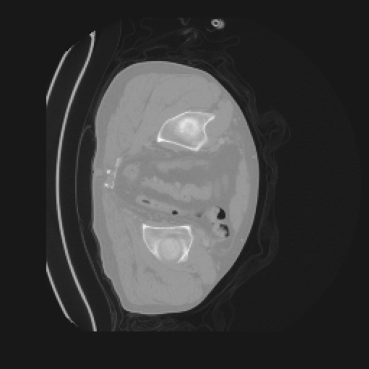}
    \end{minipage}%
      \begin{minipage}{0.15\textwidth}
        \centering
        \includegraphics[width=\textwidth]{ 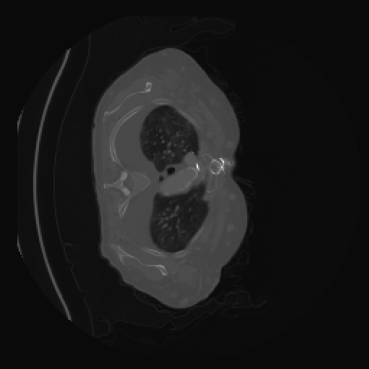}
    \end{minipage}%
      \begin{minipage}{0.15\textwidth}
        \centering
         \includegraphics[width=\textwidth]{ 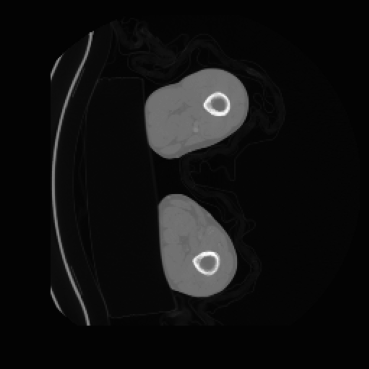}
    \end{minipage}%

        \begin{minipage}{0.15\textwidth}
        \centering
      \includegraphics[width=\textwidth]{ 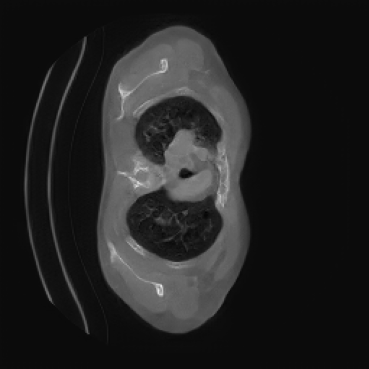}
    \end{minipage}%
    \begin{minipage}{0.15\textwidth}
        \centering
        \includegraphics[width=\textwidth]{ 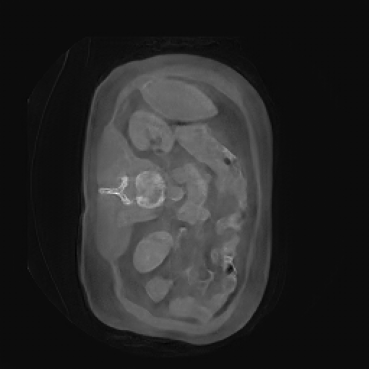} 
    \end{minipage}%
    \begin{minipage}{0.15\textwidth}
        \centering
        \includegraphics[width=\textwidth]{ 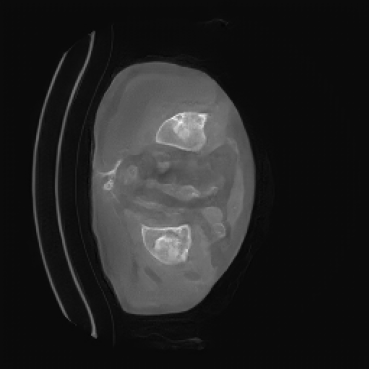}
    \end{minipage}%
      \begin{minipage}{0.15\textwidth}
        \centering
        \includegraphics[width=\textwidth]{ 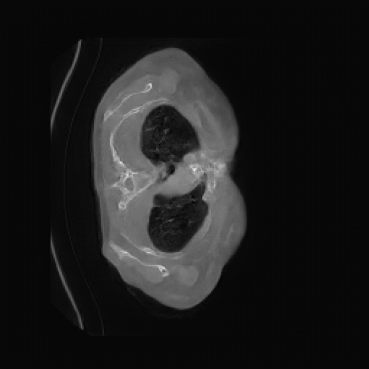}
    \end{minipage}%
      \begin{minipage}{0.15\textwidth}
        \centering
         \includegraphics[width=\textwidth]{ 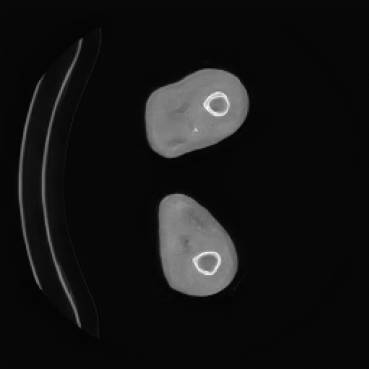}
    \end{minipage}%

    \caption{Examples of semantic map (first row), original 2D MR image slices (second row) and 2D synthetic image slices (third row) generated by Med-LSDM}
    \label{autopet_res}
\end{figure}

The motivation for this work arises from the need to address these limitations in the synthesis of 3D medical images. Current techniques do not produce high-resolution 3D images that preserve spatial coherence and capture the full complexity of medical data, as shown in Figure \ref{autopet_res}, \ref{fig:consistency}. The development of advanced 3D generative models tailored specifically for medical imaging could not only improve the quality of synthetic medical images but also help alleviate data scarcity, enhance privacy protection, reduce computational requirement and ultimately advance the use of deep learning in healthcare applications.

\section{Related Work}

\paragraph{Image-to-Image Trasnlation.}

Image-to-image translation is a domain adaptation problem, in which an image is translated from a source domain to a target domain.  A foundational work in this field is Pix2Pix \cite{isola2018imagetoimagetranslationconditionaladversarial}, which introduced conditional adversarial networks for image-to-image translation. Pix2PixHD \cite{wang2018pix2pixHD} extended this approach by incorporating a multi-scale generator and discriminator, improving image resolution and stability. VQ-GAN \cite{esser2021tamingtransformershighresolutionimage} was proposed for high-fidelity image-to-image translation by learning a discrete latent representation that preserves spatial structures while enabling efficient generation.
MedGAN \cite{Armanious_2020} is a widely cited end-to-end medical image-to-image translation framework that leverages adversarial and style-transfer losses along with the CasNet architecture to achieve high-quality results across various tasks such as PET-CT translation and MR motion correction.

DDIBs \cite{su2023dualdiffusionimplicitbridges} address image-to-image translation tasks by leveraging the inherent optimal transport properties of diffusion models, achieving notable success in applications such as stroke-to-image translation and color conversion. However, these approaches require inputs in the form of stroke drawings or natural photographs. Semantic image synthesis is a subset of image-to-image translation, where the input is a discrete semantic map, rather than continuous pixel values. This problem can not be solved well by conventional image-to-image translation methods. Therefore, specific methods for semantic image synthesis are needed.

\paragraph{Semantic Image Synthesis.}
Semantic Image Synthesis generates images from semantic maps, where each region is labeled with a specific class. This process is challenging, as it requires maintaining both the structural accuracy defined by the semantic map and the fine-grained realism of the generated image. This technique has broad applications, including virtual reality and gaming (creating realistic environments from semantic layouts), autonomous driving (generating training data from street scenes), and content creation (helping artists transform rough sketches into detailed images).

A significant advancement came with Spatially-Adaptive Normalization (SPADE) \cite{DBLP:journals/corr/abs-1903-07291}, which conditions the generation process on semantic maps at the pixel level, ensuring better control over structure and realism. SEAN (Semantic Region-Adaptive Normalization) \cite{Zhu_2020} further enhanced this approach by enabling fine-grained control over different regions, allowing specific modulation of textures and styles within localized areas.
While not originally designed for semantic image synthesis, StyleGAN2 \cite{karras2020analyzingimprovingimagequality} introduced advanced latent space manipulation, allowing fine control over attributes such as pose, lighting, and expression, making it adaptable for semantic editing. USIS \cite{eskandar2021usisunsupervisedsemanticimage} introduces a novel unpaired semantic image synthesis framework that reduces the need for large paired datasets. It employs a SPADE generator with a self-supervised segmentation loss to synthesize images with clear semantic structure and uses a wavelet-based discriminator to retain high-frequency details while matching realistic texture and color distributions. The method demonstrates strong performance across several challenging datasets.

More recently, Diffusion Models have emerged as a robust alternative to GANs. These models generate images by progressively refining noise through a diffusion process. Conditional Diffusion Models, such as Guided Diffusion \cite{ho2020denoisingdiffusionprobabilisticmodels}, leverage semantic maps to improve image realism and stability. Unlike GANs, diffusion models suffer less from training instability and mode collapse, making them highly suitable for applications in medical imaging and complex scene generation. Wang et al. proposed Semantic Diffusion Model \cite{wang2022semanticimagesynthesisdiffusion}, which proposes a unique network structure that processes the semantic layout and noisy image independently, by feeding the noisy image to the encoder and embedding the semantic layout into the decoder through multi-layer spatially-adaptive normalization operators \cite{DBLP:journals/corr/abs-1903-07291}. Stable Diffusion \cite{rombach2022highresolutionimagesynthesislatent} is a latent diffusion model that enables high-quality and efficient image synthesis by performing denoising in a learned latent space, making it a powerful alternative to traditional GAN-based methods. The state-of-the-art methods are trained on natural images, which don't work well in medical domain due to big domain gap.

\paragraph{Medical Image Synthesis via Generative Models.}
The role of DL in medical imaging is steadily increasing. A prototypical problem that can be solved by DL involves the segmentation of medical image. The reverse action, i.e., the semantic synthesis of medical images from semantic map inputs, is less often explored but holds massive potential. Synthetic data can also be used to pre-train a DL-model to improve performance, since the medical data is usually hard and expensive to obtain. 

Among these generative models, GANs have become one of the most widely utilized methods in medical image synthesis \cite{goodfellow2014generative, Yi_2019}. Notable research includes the work of A.B. Qasim et al. \cite{qasim2021redganattackingclassimbalance}, who utilized the SPADE conditional generative network \cite{DBLP:journals/corr/abs-1903-07291} to generate new images based on pre-existing masks. Their approach concentrated on 2D slice-wise synthesis of brain images, aiming to retain semantic information essential for improving segmentation tasks. However, GANs are not without their limitations. They often encounter issues such as unstable training, mode collapse, and vanishing gradients. Additionally, many existing generative techniques are primarily focused on two-dimensional (2D) images, limiting their effectiveness in medical imaging, where three-dimensional (3D) data is often crucial. A common workaround is to generate 2D slices of an organ and stack them to create a 3D volume, but this approach can lead to spatial inconsistencies and fails to capture the full 3D contextual information that is critical in medical imaging.

Studies include those by Nicholas Konz et al. \cite{konz2024anatomicallycontrollablemedicalimagegeneration}, who have proposed the SegGuidedDiff model, which employed the DDPM \cite{ho2020denoisingdiffusionprobabilisticmodels} for synthesizing new images from existing masks, focusing on 2D slice-wise image synthesis to preserve semantic information for segmentation tasks. 
Although these studies have focused mainly on 2D medical image slices, the most important diagnostic imaging modalities in modern medicine, such as magnetic resonance tomography (MRT) or computed tomography (CT), yield 3D data. Thus, methods to generate synthetic 3D data are critically needed.
Khader et al. \cite{khader2023medicaldiffusiondenoisingdiffusion} employed a diffusion model to generate 3D medical images. Their approach generates these images unconditionally, meaning that no semantic guidance or additional information is used during the generation process. Although this method demonstrates the application of diffusion models to the 3D medical imaging domain, the lack of conditional guidance limits the ability of image synthesis with controlled content.
On the other hand, Med-DDPM \cite{Dorjsembe_2024} introduced a more refined approach by incorporating a conditional diffusion model specifically for 3D brain MRI synthesis. This model provides control over the generated output through conditional information, which improves the generation process by guiding it with semantic inputs. However, in Med-DDPM, the conditioning information is added at the initial noise stage, which can lead to less precise integration of semantic guidance throughout the generation process. In this work, we aim to advance conditional mechanisms to more effectively accommodate complex and diverse anatomical structures and pathological variations. Furthermore, we investigate the integration of semantic maps at various stages of the diffusion process to enhance controllability and synthesis accuracy in 3D medical image generation. Specifically, our Med-LSTM model adopts the classical SPADE approach to incorporate semantic information during the denoising process.

\section{Methodology}
\label{sec:headings}

\begin{figure}[ht]
	\centering
	\includegraphics[width=1.0\linewidth]{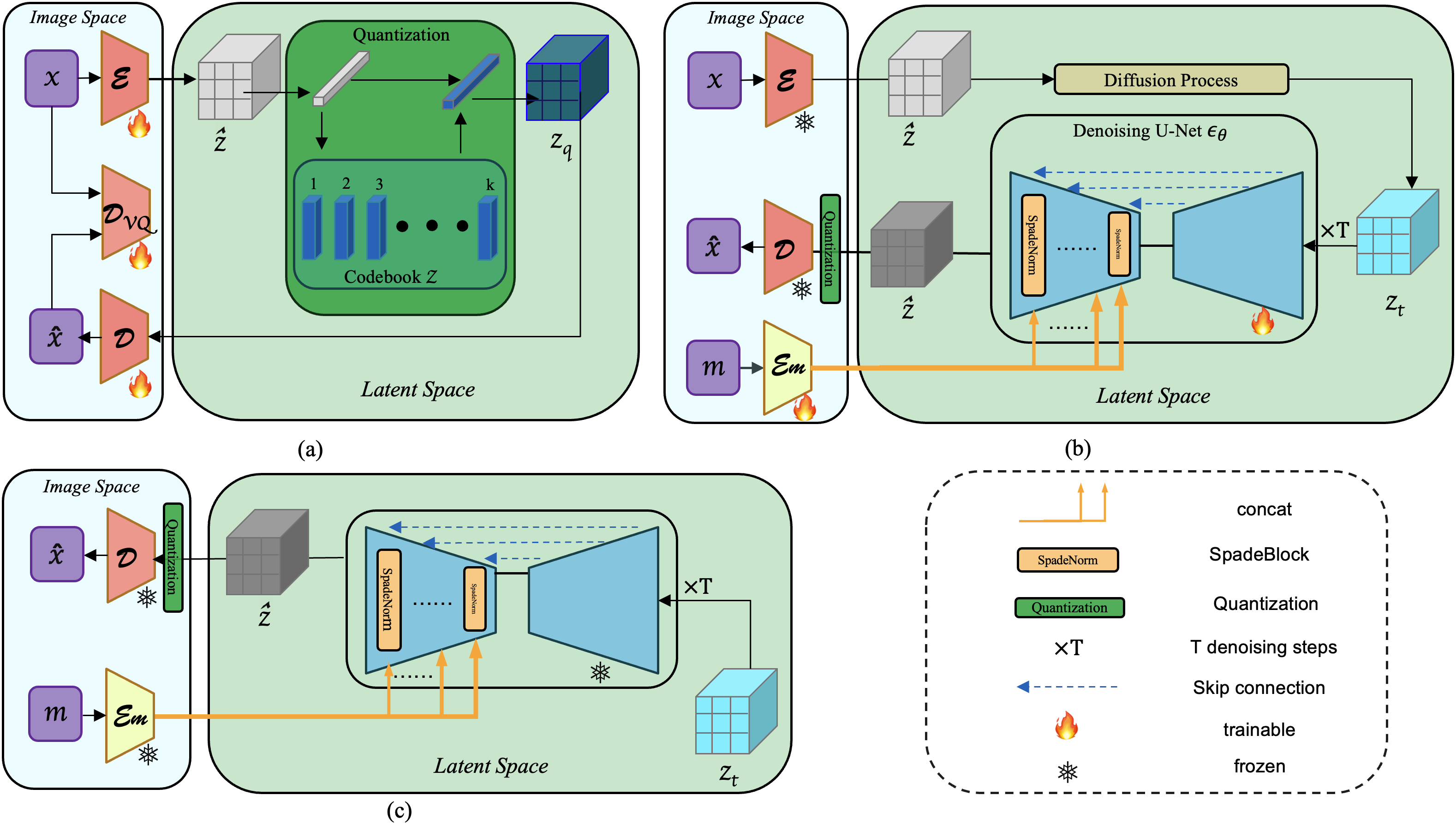} 
	\caption{Overview of the proposed Med-LSDM Framework.
(a) Training of the VQ-GAN for mapping between voxel space and latent space. (b) Training of the Semantic Diffusion Model conditioned on semantic map and input latent code. (c) Inference stage with input noise and semantic map for guided denoising in latent space.}
	\label{fig:medlsdm}
\end{figure}

In this section, we propose the Med-LSDM (Latent Semantic Diffusion Model) architecture in Figure \ref{fig:medlsdm} for 3D Semantic Medical Image Synthesis. First, we present the whole architecture, then we introduce the 3D VQ-GAN \cite{esser2021tamingtransformershighresolutionimage} and Semantic Diffusion model in detail. To the best of our knowledge, it is the first work to generate 3D medical images from semantic maps using latent diffusion.

\subsection{Med-LSDM architecture}
The architecture consists of two components: a 3D VQ-GAN and a 3D Semantic Diffusion Model, which works on the latent space of the VQ-GAN. The two components are trained separately. First, the 3D VQ-GAN is trained independently to learn a compact and discrete latent representation of the input image ${x}$. The image is encoded by the VQ-GAN encoder $\mathcal{E}$ into a latent feature $\hat{z}$, which is subsequently quantized into a discrete code ${z_q}$. A decoder $\mathcal{D}$ reconstructs the output $\hat{x}$ from the quantized latent code ${z}_q$, similar to the decoding process in an autoencoder. The discriminator \emph{\(\mathcal{D}_{VQ}\)} is employed to distinguish between real images ${x}$ and reconstructed images $\hat{x}$, thereby facilitating the learning of a discrete and expressive latent space for ${x}$. It is worth noting that the VQ-GAN model is trained on an unlabeled dataset, as it does not require paired image-label supervision, but instead learns to reconstruct input data through adversarial and perceptual objectives. Once the VQ-GAN is trained, its parameters are frozen. 

Subsequently, the 3D Semantic Diffusion Model is trained on the fixed latent space. During this phase, the semantic map ${m}$ is processed through a semantic map encoder $\mathcal{E}_{m}$ and integrated into the denoising steps to guide the diffusion process with semantic information. The input image ${x}$ is encoded by the frozen VQ-GAN encoder $\mathcal{E}$ into a latent feature $\hat{{z}}$, which undergoes a forward diffusion process, progressively adding noise over time to generate noisy intermediate states ${z}_t$. The Denoising U-Net, denoted by \( \epsilon_{\theta} \), attempts to reverse this process in ${T}$ steps, removing noise in a stepwise manner to restore the noisy latent representation to its clean state. The U-Net utilizes \textit{SPADE} \cite{DBLP:journals/corr/abs-1903-07291} normalization to condition the denoising process on the semantic map ${m}$, enabling the model to preserve semantic information throughout. Finally, the denoised latent code is quantized based on the frozen codebook \emph{$\mathcal{Z}$} and passed through the frozen VQ-GAN decoder $\mathcal{D}$ to reconstruct the output image $\hat{{x}}$, which closely resembles the original input ${x}$. This approach leverages the efficiency of working in a compressed latent space and the effectiveness of diffusion models for generating diverse and high-quality images, making it particularly suitable for tasks like semantic image synthesis and image-to-image translation. In this phase, the VQ-GAN is kept frozen, and only the semantic map encoder $\mathcal{E}_m$ and the Denoising U-Net \( \epsilon_{\theta} \) are trained using paired data $({x}, {m})$, where ${m}$ denotes the semantic map corresponding to the image ${x}$.

During inference, the model generates 3D images solely based on a random noise vector and a given semantic map. Specifically, a noise sample ${z}_t$ is sampled from a standard Gaussian distribution, representing the highly corrupted latent state. The semantic map ${m}$ is processed through the same semantic map encoder $\mathcal{E}_{m}$. Conditioned on these semantic features, the frozen Semantic Diffusion Model progressively denoises ${z}_T$ over multiple steps, ultimately recovering a clean latent code $\hat{{z}}$. Finally, the frozen VQ-GAN decoder $\mathcal{D}$ maps the denoised latent code $\hat{{z}}$ back into the image space, producing the synthesized 3D image. Notably, the ground-truth image ${x}$ is not required during inference.

\subsection{VQ-GAN}
The VQ-GAN architecture proposed by Esser et al. \cite{esser2021tamingtransformershighresolutionimage} is used to connect the voxel space and the compressed latent space. The VQ-GAN model is a GAN designed to generate high-quality images using a codebook-based vector quantization (VQ) mechanism. The architecture comprise six components: an Encoder \emph{$\mathcal{E}$}, a Decoder \emph{$\mathcal{D}$}, a learned Codebook \emph{$\mathcal{Z}$}, Discriminators  \emph{\(\mathcal{D}_{VQ}\)} (including a 2D discriminator \emph{\(\mathcal{D}_{2D}\)} and a 3D discriminator \emph{\(\mathcal{D}_{3D}\)}), and a perceptual model. The 3D discriminator \emph{\(\mathcal{D}_{3D}\)} takes the output 3D images $\hat{{x}}$ and input 3D images ${{x}}$ as input, while the 2D discriminator \emph{\(\mathcal{D}_{2D}\)} receives randomly sampled 2D slices from $\hat{{x}}$ and ${{x}}$. The VQ mechanism enables any image \( x \in \mathbb{R}^{H \times W \times L \times C} \) to be represented as a spatial collection of codebook entries \( z_q \in \mathbb{R}^{(H/t) \times (W/t) \times (L/t) \times n_z} \), where \( H, W, L \) denote the spatial scale of 3D input image, \( C\) denotes number of channels,   \( t \) denotes the factor of spatial compression, \( n_z \) denotes the dimension of the latent codes  
\begin{equation}
\hat{z} = \emph{$\mathcal{E}$}(x) \in \mathbb{R}^{(H/t) \times (W/t) \times (L/t) \times n_z} \ \,
\end{equation}
\( \hat{z} \) is then quantized element-wise to the nearest entry in the  codebook \(\emph{$\mathcal{Z}$} = \{z_k\}_{k=1}^K \in \mathbb{R}^{n_z} \) :

\[
z_q = q(\hat{z}) := \arg \min_{z_k \in \emph{$\mathcal{Z}$}} \|\hat{z}_{h,w,l} - z_k\| \quad \text{for each } \hat{z}_{h,w,l} \in \mathbb{R}^{n_z} .
\]

Quantization in VQ-GAN discretizes the continuous latent space by mapping encoder outputs to a learned codebook. This regularizes the latent space, improves the training stability, enhances the quality of reconstruction, and facilitates the integration with token-based generative models.

The quantized latent representation $ z_q $ is then passed through the decoder \emph{\(\mathcal{D}\)} to reconstruct the image:

\begin{equation}
\hat{x} = \emph{$\mathcal{D}$}(z_q) = \emph{$\mathcal{D}$}(q(\emph{$\mathcal{E}$}(x))) .
\end{equation}

The quantization operation \( q(\cdot) \) is inherently non-differentiable, posing a challenge for back-propagation. To address this, Esser et al. \cite{esser2021tamingtransformershighresolutionimage} employ a straight-through gradient estimator, which effectively bypasses the quantization step by copying gradients from the decoder directly to the encoder. This allows the model to be trained end-to-end using the loss function:

\begin{equation}
\mathcal{L}_{\text{VQ}}(\emph{$\mathcal{E}$}, \emph{$\mathcal{D}$}, \emph{$\mathcal{Z}$}) = \|x - \hat{x}\|^2_2 + \|\text{sg}[\emph{$\mathcal{E}$}(x)] - z_q\|^2_2 + \|\text{sg}[z_q] - \emph{$\mathcal{E}$}(x)\|^2_2 .
\end{equation}

Here, \( \mathcal{L}_{\text{rec}} = \|x - \hat{x}\|^2_2 \) is the reconstruction loss, \( \text{sg}[\cdot] \) denotes the stop-gradient operation, and \( \|\text{sg}[z_q] - \emph{$\mathcal{E}$}(x)\|^2_2 \) is the commitment loss, which encourages the encoder's output to remain close to the quantized values.

To maintain high perceptual quality while pushing the limits of compression, we apply VQ-GAN. The adversarial loss and perceptual loss are defined as:

\begin{equation}
\mathcal{L}_{\text{GAN}}(\{\emph{$\mathcal{E}$}, \emph{$\mathcal{D}$}, \emph{$\mathcal{Z}$}\}, \emph{$\emph{\(\mathcal{D}_{VQ}\)}$}) = \log \emph{\(\mathcal{D}_{VQ}\)}(x) + \log(1 - \emph{\(\mathcal{D}_{VQ}\)}(\hat{x})) ,
\end{equation}

\begin{equation}
\mathcal{L}_{\text{perc}}(\emph{$\mathcal{E}$}, \emph{$\mathcal{D}$}, \emph{$\mathcal{Z}$}) = \sum_{l} \left\| \phi_l(x) - \phi_l(\hat{x}) \right\|_2^2 ,
\end{equation}

where, $\phi_l$ is the feature mapping function of the $l$-th layer of a pre-trained network (VGG).

The complete objective for learning the optimal compression model \( \mathcal{Q}^* = \{\mathcal{E}^*, \emph{$\mathcal{D}$}^*, \emph{$\mathcal{Z}$}^*, \emph{\(\mathcal{D}_{VQ}\)}^* \} \) is formulated as:

\begin{equation}
\mathcal{Q}^* = \arg \min_{\mathcal{E}, \mathcal{D}, \mathcal{Z}} \max_{\mathcal{D}_{VQ}} \mathbb{E}_{x \sim p(x)} \left[ \mathcal{L}_{\text{VQ}}(\mathcal{E}, \mathcal{D}, \mathcal{Z}) + \lambda \mathcal{L}_{\text{GAN}}(\{\mathcal{E}, \mathcal{D}, \mathcal{Z}\}, \emph{\(\mathcal{D}_{VQ}\)}) + \alpha \mathcal{L}_{\text{perc}}(\mathcal{E}, \mathcal{D}, \mathcal{Z}) \right],
\end{equation}

where the dynamic weighting factor \( \lambda \) is determined by the relative gradients of the reconstruction loss \( \mathcal{L}_{\text{rec}} \) and the adversarial loss \( \mathcal{L}_{\text{GAN}} \):

\begin{equation}
\lambda = \frac{\left\| \nabla_{G^L} \mathcal{L}_{\text{rec}} \right\|_2}{\left\| \nabla_{G^L} \mathcal{L}_{\text{GAN}} \right\|_2 + \delta}
\end{equation}

Here, \( \nabla_{G^L} \mathcal{L}_{\text{rec}} \) and \( \nabla_{G^L} \mathcal{L}_{\text{GAN}} \) denote the gradients of the perceptual reconstruction loss and the adversarial loss with respect to the last layer \( G^L \) of the decoder. The symbol \( \| \cdot \|_2 \) represents the \( \ell_2 \)-norm, and \( \delta = 10^{-6} \) is a small constant added to ensure numerical stability.

\subsection{Latent Semantic Diffusion Model}

The Semantic Diffusion Model (SDM) is a conditional denoising diffusion probabilistic model designed for high-fidelity 3D medical image synthesis. It leverages the standard denoising diffusion framework, where the generation process involves gradually refining random noise conditioned on a semantic map. To align with conventional notation, we denote the input latent space representation as \( {z}_0 \) instead of \( \hat{{z}} \). The model is trained to approximate the true conditional distribution \( q({z}_0 \mid m) \) by learning a reverse process \( p_\theta({z}_0 \mid m) \), where \( {z}_0 \) lies in the latent space, \( m \) is the semantic condition, and \( \theta \) are the model parameters.

The forward process \( q(z_{1:T} \mid z_0) \) gradually adds Gaussian noise to the data \( {z}_0 \), producing a sequence of increasingly noisy latent variables \( {z}_1, {z}_2, ... , {z}_T  \). The reverse process \( p_\theta(z_{0:T} \mid m) \) defines a Markov chain that denoises step-by-step, starting from an initial Gaussian distribution \( p(z_T) = \mathcal{N}(0, I) \), and aims to recover the clean sample \( z_0 \) conditioned on \( m \). Each subsequent state in the chain is defined by learned Gaussian transitions, encapsulated in the following formulation:
\begin{equation}
    p_\theta(z_{0:T}|m) = p(z_T) \prod_{t=1}^T p_\theta(z_{t-1}|z_t, m),
\end{equation}
\begin{equation}
    p_\theta(z_{t-1}|z_t, m) = \mathcal{N}(z_{t-1}; \mu_\theta(z_t, m, t), \Sigma_\theta(z_t, m, t)).
\end{equation}
Here, $\mu_\theta(z_t, m, t)$ and $\Sigma_\theta(z_t, m, t)$ represent the mean and variance functions parameterized by neural networks, which learn to map the noisy inputs back to clean data through the conditioning variable $m$.

Conversely, the forward process $q(z_{1:T}|z_0)$ involves the gradual addition of Gaussian noise to the data $z_0$ according to a predefined variance schedule $\beta_1, \beta_2, \ldots, \beta_T$.  that controls the rate of noise addition across \( T \) diffusion steps. Here we chose cosine-increasing schedule, where smaller \( \beta_t \) values at earlier steps preserve more of the original data, and larger \( \beta_t \) values at later steps allow more noise and flexibility. This can be expressed as:
\begin{equation}
    q(z_t|z_{t-1}) = \mathcal{N}(z_t; \sqrt{1 - \beta_t} z_{t-1}, \beta_t I).
\end{equation}
To simplify notation, we define $\alpha_t := \prod_{n=1}^t (1 - \beta_n)$, allowing us to reformulate the distribution as:
\begin{equation}
    q(z_t|z_0) = \mathcal{N}(z_t; \sqrt{\alpha_t} z_0, (1 - \alpha_t) I).
\end{equation}
This forward process progressively corrupts the data, enabling the model to learn the reverse process that denoises it.



To align with the standard conditional DDPM framework, we model the denoising process by predicting the noise added to the latent variable \( z_0 \) during the forward process. Specifically, the forward diffusion adds Gaussian noise to the latent variable \( z_0 \), resulting in a noisy sample \( z_t \), formulated as:
\begin{equation}
    z_t = \sqrt{\alpha_t} z_0 + \sqrt{1 - \alpha_t} \epsilon, \quad \epsilon \sim \mathcal{N}(0, I),
\end{equation}

During training, the model learns a noise prediction network \( \epsilon_\theta(z_t, t, m) \) that estimates the true noise \( \epsilon \) given the noisy sample \( z_t \), timestep \( t \), and semantic condition \( m \). Following the terminology introduced by Ho et al.~\cite{ho2020denoisingdiffusionprobabilisticmodels}, we denote the noise prediction objective as \(\mathcal{L}_{\text{simple}}\), which is used in the original denoising diffusion probabilistic model. Instead of computing multiple KL divergence terms, the simplified loss directly regresses the noise added to the data using an L2 objective:
\begin{equation}
    \mathcal{L}_{\text{simple}} = \mathbb{E}_{z_0, \epsilon, t} \left[ \left\| \epsilon - \epsilon_\theta(z_t, t, m) \right\|^2 \right].
\end{equation}
This formulation avoids reconstructing \( z_0 \) directly and instead focuses on accurately denoising through iterative refinement steps, which is computationally stable and aligns with the widely used DDPM paradigm.

The SDM is based on a U-Net architecture, which consists of multiple Resblocks. The feature of the noisy image is encoded through stacked semantic diffusion encoder residual blocks (Encoder Resblocks) and attention blocks. Figure \ref{blocks} illustrates the structure of the Econder Resblocks, which comprises convolution layers, SiLU activation and group normalization. SiLU \cite{ramachandran2017searchingactivationfunctions} often performs better than ReLU \cite{nair2010rectified} in deeper networks. To enable the network to estimate noise at various time steps $t$, the Encoder Resblock involves scaling and shifting the intermediate activations using a learnable weight \( w(t) \in \mathbb{R}^{1 \times 1 \times 1 \times C} \) and bias \( b(t) \in \mathbb{R}^{1 \times 1 \times 1 \times C} \), formulated as:

 \begin{figure}[h]
	\centering
	\includegraphics[scale=.52]{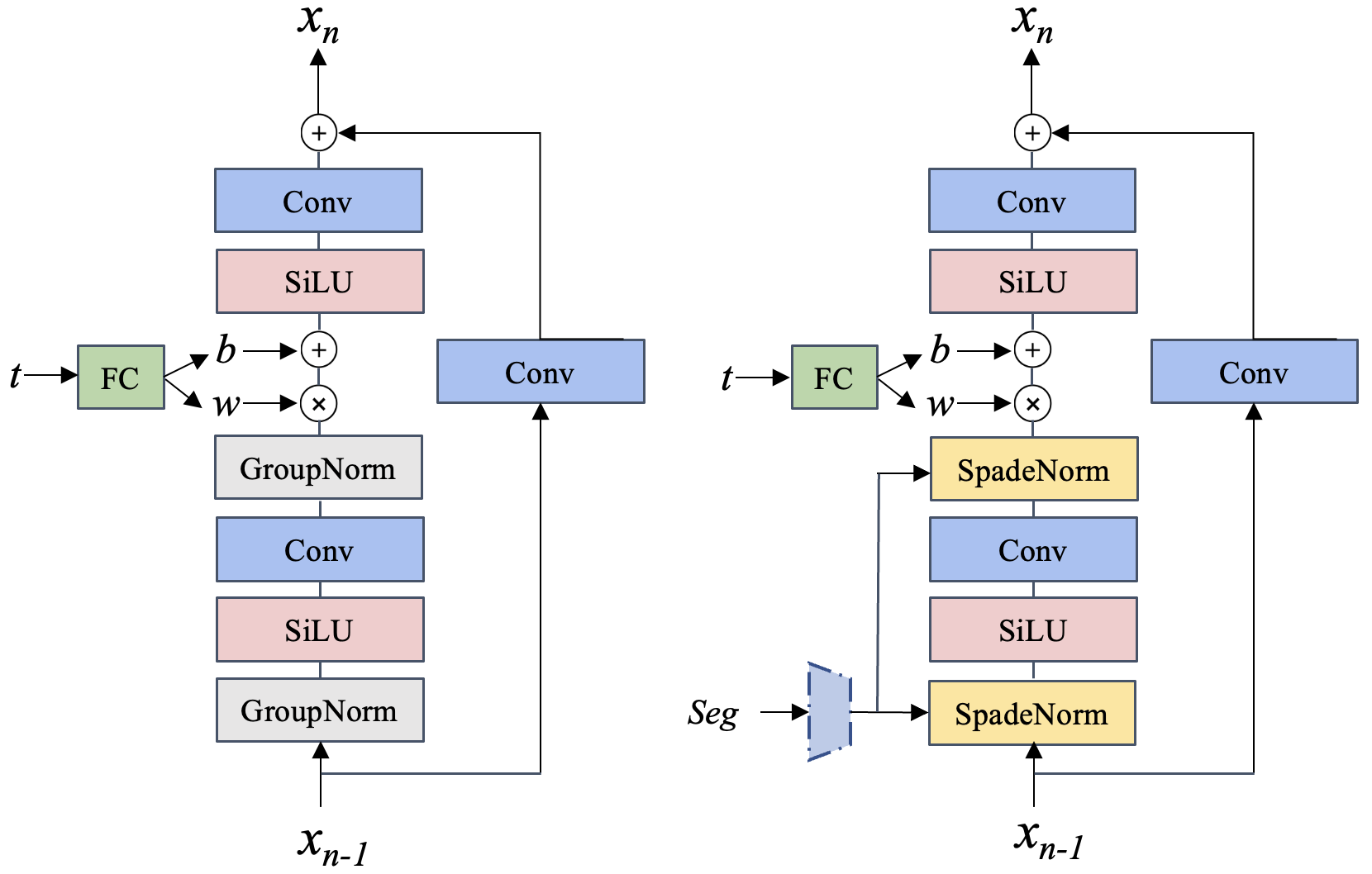}
	\caption{Encoder and Decoder Resblock of denosing U-Net}
	\label{blocks}
\end{figure}

\begin{equation}
f_{i+1} = w(t) \cdot f_i + b(t),
\end{equation}

where \( f_i, f_{i+1} \in \mathbb{R}^{H \times W \times L \times C} \) represent the input and output features respectively.

In 3D medical imaging such as CT or MRI scans, the data comprises three spatial dimensions (height, width, and depth). 3D volumes are often processed as sequences of 2D slices along the depth axis, which resemble temporal sequences in structure. To effectively model both inter-slice relationships and intra-slice spatial features, our Med-LSDM model incorporates two types of attention mechanisms: cross-slice attention and spatial attention. The cross-slice attention module captures dependencies across adjacent slices in the volumetric data. This is crucial for maintaining anatomical consistency and modeling long-range context across the depth of the scan, which would otherwise be challenging using purely convolutional approaches. In contrast, the spatial attention module operates within each individual slice, focusing on salient pixel-level features and capturing global spatial context to enhance local structure representation.
By combining cross-slice and spatial attention, our model effectively leverages both inter-slice coherence and intra-slice detail, enabling the extraction of rich and consistent features from complex 3D medical images.

Our model integrates the semantic map into the decoder of the denoising network independently to enhance the 3D image generation process. Previous conditional diffusion models \cite{konz2024anatomicallycontrollablemedicalimagegeneration, Dorjsembe_2024}, which directly concatenate condition information with noisy 3D image as input, fail to fully exploit semantic information, lead to generated images with lower quality and weaker semantic relevance. To address this limitation, we apply the semantic diffusion decoder Resblock (see Figure \ref{blocks}) to embed the semantic map into the decoder of the denoising network in a multi-layer spatially-adaptive manner. We employ SPADE \cite{DBLP:journals/corr/abs-1903-07291} instead of group normalization to handle 3D images and semantic maps. The SPADE injects the semantic label map into the denoising stream by adjusting the feature maps using spatially adaptive, learnable transformations, which is mathematically expressed as:

\begin{equation}
f_{i+1} = \gamma_i(x) \cdot \text{Norm}(f_i) + \beta_i(x),
\end{equation}

where $f_i$ and $f_{i+1}$ denote the input and output features of the SPADE module. $\text{Norm}(\cdot)$ refers to the parameter-free group normalization. $\gamma_i(x)$ and $\beta_i(x)$ are the spatially-adaptive weight and bias terms, respectively, which are learned from the semantic map. It is worth mentioning that in order to reduce the information loss of the down sampling operation of the semantic map, the semantic map is firstly processed by a semantic map encoder $\mathcal{E}_{m}$ before input to the SPADE block.





\section{Experiments}

We evaluate our method on three benchmark datasets: AutoPET (CT)~\cite{gatidis2022whole}, SynthRAD2023 (paired CT and MR)~\cite{Thummerer_2023}, and Duke Breast Dataset (MR)~\cite{breastcancer} as shown in Tabel \ref{tab:dataset_properties}. The AutoPET dataset provides 3D medical images along with their corresponding semantic maps, enabling direct training and evaluation of both the VQ-GAN and latent diffusion model. The Duke Breast dataset includes a subset of 100 scans that are annotated with semantic maps. To train the VQ-GAN, we utilize additional unlabeled scans outside of this 100-scan subset. Among the labeled scans, 90 are used to train the latent diffusion model, and the remaining 10 scans serve as the test set.  In contrast, the SynthRAD2023 dataset only includes paired CT and MR images without associated semantic annotations. We To address this, we leverage the open-source medical image segmentation tool \textit{TotalSegmentator}~\cite{Wasserthal_2023} to generate semantic maps from the CT images. These generated CT semantic maps, together with the corresponding MR images, are then used to train the Semantic Diffusion Model, enabling us to evaluate the cross-domain generalization capability of our framework. The VQ-GAN is trained separately using the image data from each dataset without requiring any semantic maps. Specifically, we train separate VQ-GAN models for each dataset: one model using CT images from the AutoPET training dataset, another using MR images from the Duke Breast training dataset, and a third using MR images from SynthRAD2023 training dataset. Since the VQ-GAN operates in a self-reconstruction manner, it does not rely on semantic maps and can be trained in a self-supervised manner. This allows each model to learn a compact and discrete latent space that captures high-level image features specific to its respective imaging modality and dataset characteristics, which are later used in the semantic diffusion stage.

\begin{table}[h!]
\centering
\caption{Comparison of dataset properties used in our experiments.}
\label{tab:dataset_properties}
\begin{tabular}{c|c|c|c|c}
 \toprule
   Dataset       & 3D Scans (Train/Test) & Resolution & Voxel Size (mm) & Semantic Classes\\ 
 \midrule
   AutoPET (CT)~\cite{gatidis2022whole}         
     & 912/102 & 304 × 400 × 400 & 2×2×3 & 37 \\ 
     
   SynthRAD2023 (CT/MR)~\cite{Thummerer_2023}   
     & 162/18  & 32 × 192 × 360  & 1×1×1 & 31 \\ 
     
   Duke Breast (MR)~\cite{konz2024anatomicallycontrollablemedicalimagegeneration} 
     & (822 + 90)/10 
     & 60 × 512 × 512 & 2×2×3 & 3 \\ 
 \bottomrule
\end{tabular}

\vspace{1mm}
\begin{minipage}{0.9\textwidth}
\small
\textit{Note:} Resolution and voxel size values may vary across samples due to differences in scanning protocols.  
The Duke Breast dataset contains 100 labeled scans with semantic maps (90 for training, 10 for testing), and 822 additional unlabeled scans used to train the VQ-GAN.
\end{minipage}
\end{table}

We perform 3D image synthesis. Due to the limitation of GPU memory, it is impossible to use the raw 3D volumetric medical imaging data as input directly, which is more than 30 MB per sample, so we process the data to the size (256, 256, 32, 1), which means each sample consists of 32 slices and each slice has one channel and image size 256*256. We set compression factor \( t = 4 \), the input medical image \( {x} \in \mathbb{R}^{256 \times 256 \times 32 \times 1} \) is downsampled by the VQ-GAN encoder to a latent representation \( \hat{z} \in \mathbb{R}^{64 \times 64 \times 8 \times 8} \), which serves as the training input for the semantic diffusion model. Here, the dimensions correspond to \((H, W, L, C)\), where \( H \), \( W \), and \( L \) are the spatial dimensions and \( C \) is the number of latent channels. The VQ-GAN model was trained with a batch size of 2 on a single RTX A5000 GPU with 24 G memory, and the SDM model was trained with a batch size of 2 on a single RTX A6000 GPU with 48 G memory. For the optimization process in all of our experiments, we employed the Adam algorithm, with the momentum terms $\beta_1$ and $\beta_2$ set to 0.9 and 0.999, respectively, and a learning rate of 0.0003.
 
In the first phase of our approach, we train a VQ-GAN model to learn a compact and meaningful latent representation of 3D medical images. To ensure compatibility with the diffusion model, which requires input normalized between -1 and 1, we enforce this range on the latent representations. The codebook consists of 16384 discrete latent codes \( (K = 16384) \) , each represented by an 8-dimensional vector \( (n_z = 8) \) . Assuming that the vector quantization step encourages the learned codebook vectors to approximate the unquantized latent features, we estimate the feature range using the minimum and maximum values from the codebook. Applying Min-Max normalization to the unquantized latent vectors maps them to the desired range, making them suitable for training the 3D latent diffusion model. During inference, new images are synthesized by applying the reverse diffusion process to Gaussian noise, followed by quantization through the VQ-GAN’s learned codebook and reconstruction through its decoder.

To demonstrate the advantages of 3D semantic image synthesis over 2D semantic semantic image synthesis, we conducted a detailed comparison with the \textit{SegGuidedDiff} \cite{konz2024anatomicallycontrollablemedicalimagegeneration} model. SegGuidedDiff generates medical images based on 2D semantic maps, producing individual 2D slices that are later stacked to form a 3D image. However, this approach has limitations in capturing spatial continuity and coherence across slices, which is critical for representing the spatial structure of 3D medical images. As a result, inconsistencies often arise between different slices, particularly in preserving anatomical structures across the coronal and sagittal planes. The generated images exhibit varying levels of denoising, resulting in differences in brightness and sharpness across the images. This inconsistency in denoising can lead to noticeable variations in visual quality, where some regions appear clearer and others less distinct.

\begin{figure}[h]
    \centering
          \begin{minipage}{0.12\textwidth}
        \centering
        \includegraphics[width=\textwidth]{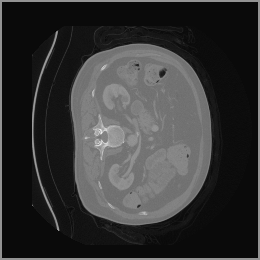}
    \end{minipage}%
    \begin{minipage}{0.12\textwidth}
        \centering
        \includegraphics[width=\textwidth]{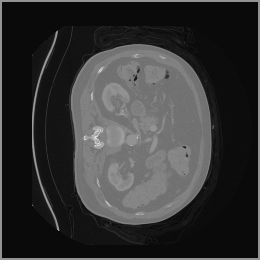} 
    \end{minipage}%
    \begin{minipage}{0.12\textwidth}
        \centering
        \includegraphics[width=\textwidth]{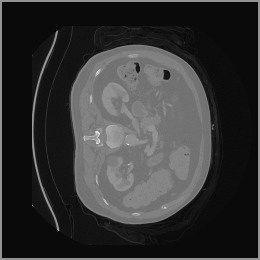}
    \end{minipage}%
      \begin{minipage}{0.12\textwidth}
        \centering
        \includegraphics[width=\textwidth]{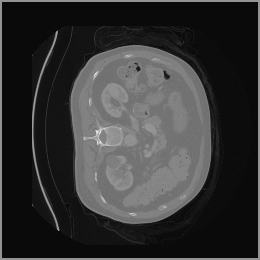}
    \end{minipage}%
      \begin{minipage}{0.12\textwidth}
        \centering
        \includegraphics[width=\textwidth]{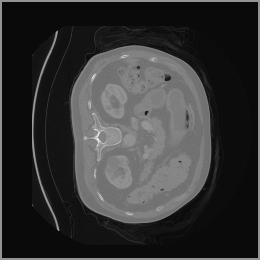}
    \end{minipage}%
      \begin{minipage}{0.12\textwidth}
        \centering
        \includegraphics[width=\textwidth]{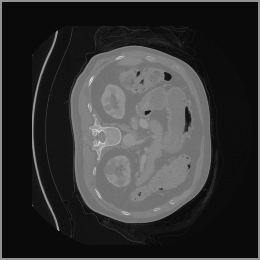}
    \end{minipage}%
      \begin{minipage}{0.12\textwidth}
        \centering
        \includegraphics[width=\textwidth]{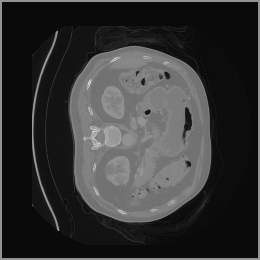}
    \end{minipage}%
      \begin{minipage}{0.12\textwidth}
        \centering
        \includegraphics[width=\textwidth]{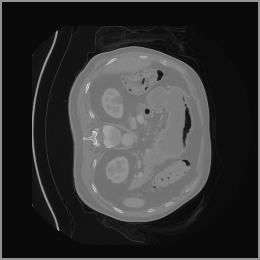}
    \end{minipage}%

            \begin{minipage}{0.12\textwidth}
        \centering
        \includegraphics[width=\textwidth]{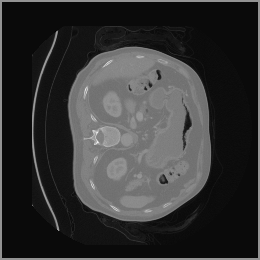}
    \end{minipage}%
    \begin{minipage}{0.12\textwidth}
        \centering
        \includegraphics[width=\textwidth]{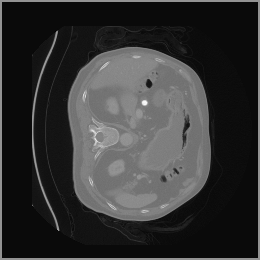} 
    \end{minipage}%
    \begin{minipage}{0.12\textwidth}
        \centering
        \includegraphics[width=\textwidth]{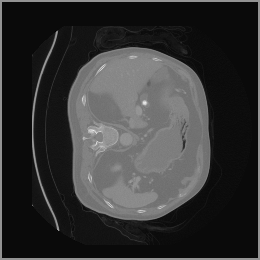}
    \end{minipage}%
      \begin{minipage}{0.12\textwidth}
        \centering
        \includegraphics[width=\textwidth]{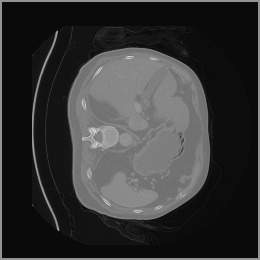}
    \end{minipage}%
      \begin{minipage}{0.12\textwidth}
        \centering
        \includegraphics[width=\textwidth]{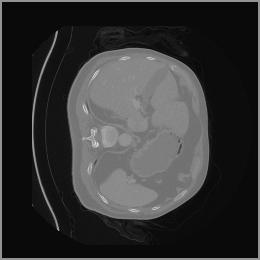}
    \end{minipage}%
      \begin{minipage}{0.12\textwidth}
        \centering
        \includegraphics[width=\textwidth]{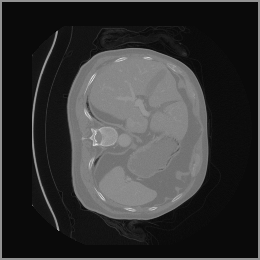}
    \end{minipage}%
      \begin{minipage}{0.12\textwidth}
        \centering
        \includegraphics[width=\textwidth]{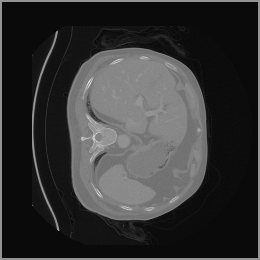}
    \end{minipage}%
      \begin{minipage}{0.12\textwidth}
        \centering
        \includegraphics[width=\textwidth]{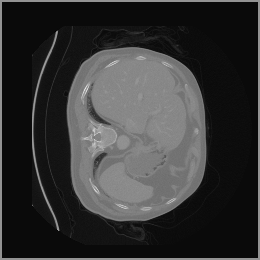}
    \end{minipage}%

          \begin{minipage}{0.12\textwidth}
        \centering
        \includegraphics[width=\textwidth]{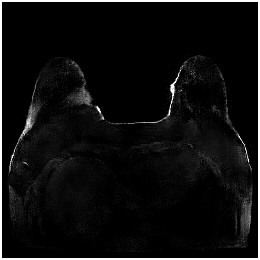}
    \end{minipage}%
    \begin{minipage}{0.12\textwidth}
        \centering
        \includegraphics[width=\textwidth]{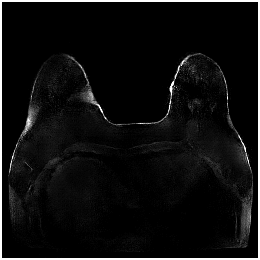} 
    \end{minipage}%
    \begin{minipage}{0.12\textwidth}
        \centering
        \includegraphics[width=\textwidth]{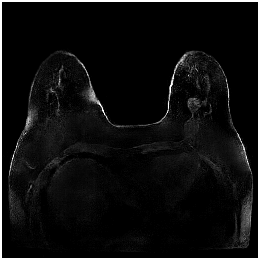}
    \end{minipage}%
      \begin{minipage}{0.12\textwidth}
        \centering
        \includegraphics[width=\textwidth]{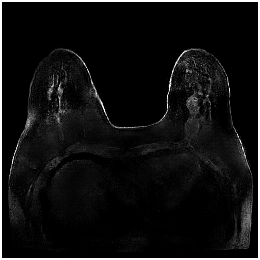}
    \end{minipage}%
      \begin{minipage}{0.12\textwidth}
        \centering
        \includegraphics[width=\textwidth]{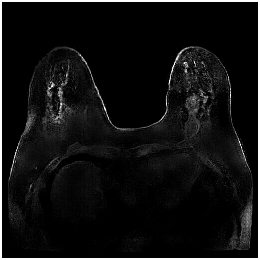}
    \end{minipage}%
      \begin{minipage}{0.12\textwidth}
        \centering
        \includegraphics[width=\textwidth]{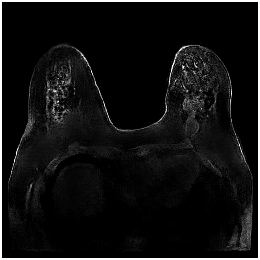}
    \end{minipage}%
      \begin{minipage}{0.12\textwidth}
        \centering
        \includegraphics[width=\textwidth]{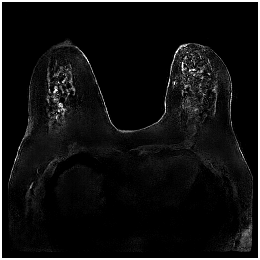}
    \end{minipage}%
      \begin{minipage}{0.12\textwidth}
        \centering
        \includegraphics[width=\textwidth]{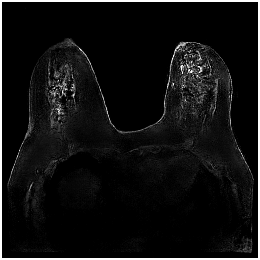}
    \end{minipage}%

            \begin{minipage}{0.12\textwidth}
        \centering
        \includegraphics[width=\textwidth]{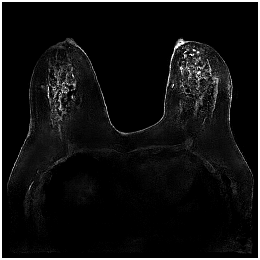}
    \end{minipage}%
    \begin{minipage}{0.12\textwidth}
        \centering
        \includegraphics[width=\textwidth]{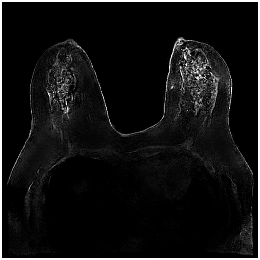} 
    \end{minipage}%
    \begin{minipage}{0.12\textwidth}
        \centering
        \includegraphics[width=\textwidth]{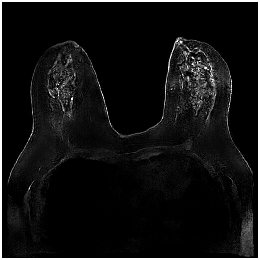}
    \end{minipage}%
      \begin{minipage}{0.12\textwidth}
        \centering
        \includegraphics[width=\textwidth]{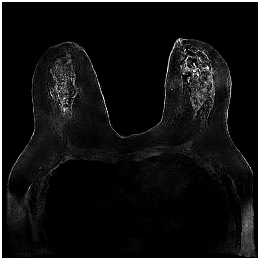}
    \end{minipage}%
      \begin{minipage}{0.12\textwidth}
        \centering
        \includegraphics[width=\textwidth]{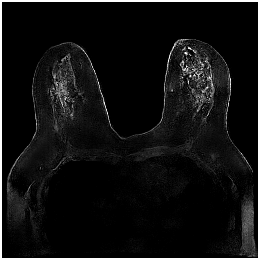}
    \end{minipage}%
      \begin{minipage}{0.12\textwidth}
        \centering
        \includegraphics[width=\textwidth]{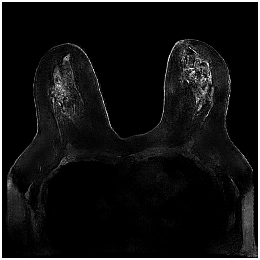}
    \end{minipage}%
      \begin{minipage}{0.12\textwidth}
        \centering
        \includegraphics[width=\textwidth]{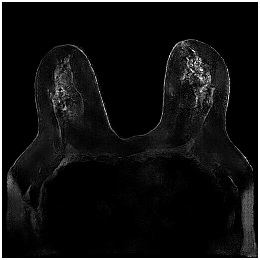}
    \end{minipage}%
      \begin{minipage}{0.12\textwidth}
        \centering
        \includegraphics[width=\textwidth]{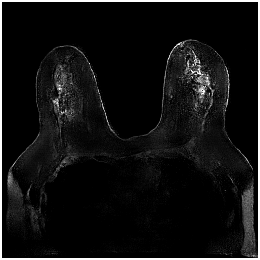}
    \end{minipage}%

    \caption{One 3D CT scan (top rows) and one MRT scan (bottom rows), which are generated by our model from the semantic maps of AutoPET CT and Duke Breast MRI dataset respectively.  Each sample consists of 32 slices. To reduce redundancy, we uniformly sampled one slice every two slices from each sample, resulting in a set of representative images for qualitative evaluation. }
    \label{fig:consistency}
\end{figure}

\subsection{Med-LSDM can Generate High Quality 3D Synthetic Images}

We evaluated the synthetic images from three aspects: 
\begin{itemize}
    \item \textbf{Inter-slice consistency} - Evaluation of  the coherence and continuity between adjacent slices in 3D volumes.
    \item \textbf{Overall image quality} - Evaluation of visual realism and quality of the generated image features. 
    \item \textbf{Faithfulness to input masks} - Measuring how well the generated images align with the provided semantic masks.
\end{itemize}

As shown in Figure \ref{fig:consistency}, the generated slices exhibit high consistency across adjacent slices. The anatomical structures and semantic patterns are smoothly and coherently preserved between slices, indicating that our model is capable of generating 3D-consistent synthetic images rather than isolated 2D slices.

We employ 3D-FID \cite{9770375} based on a 3D network called Med3D \cite{chen2019med3dtransferlearning3d} , which is pre-trained on 3DSeg-8 dataset with diverse
modalities, target organs, and pathologies to extract general medical three-dimension (3D) features for calculating the Fréchet Inception Distance (FID).

\subsection{Faithfulness Evaluation of Generated Images to Input Masks.}

\begin{figure}[ht]
    \centering
    \small
  
    \includegraphics[width=0.15\textwidth]{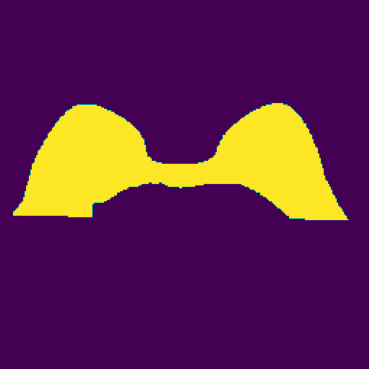} 
    \includegraphics[width=0.15\textwidth]{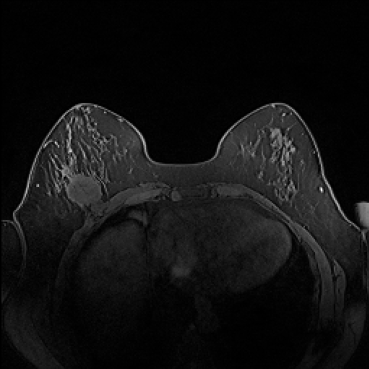} 
    \includegraphics[width=0.15\textwidth]{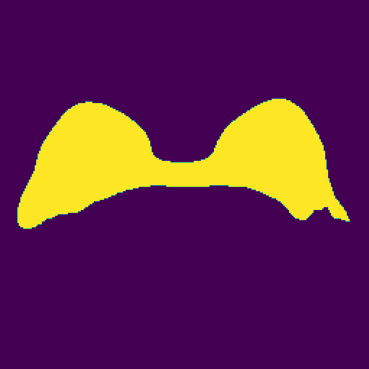}
     \includegraphics[width=0.15\textwidth]{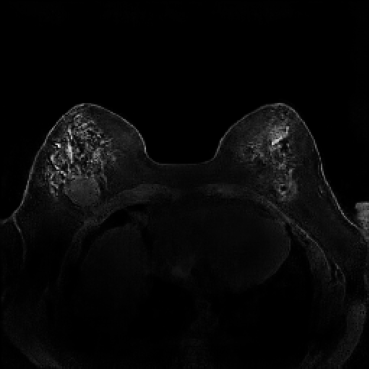} 
    \includegraphics[width=0.15\textwidth]{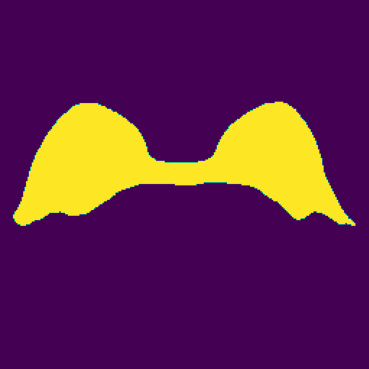} \\

   \includegraphics[width=0.15\textwidth]{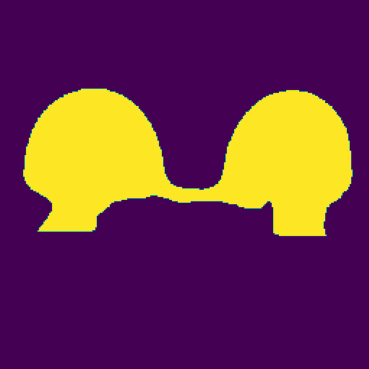} 
    \includegraphics[width=0.15\textwidth]{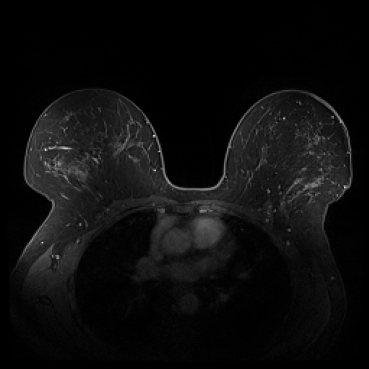} 
    \includegraphics[width=0.15\textwidth]{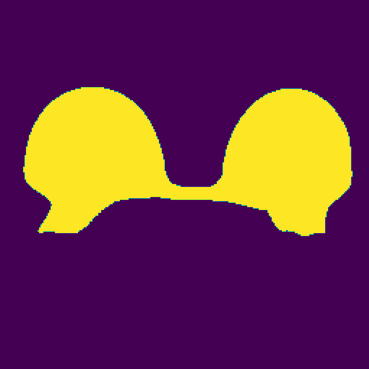}
     \includegraphics[width=0.15\textwidth]{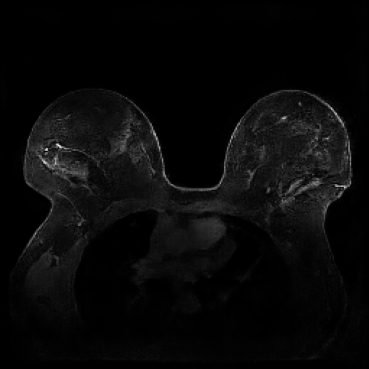} 
    \includegraphics[width=0.15\textwidth]{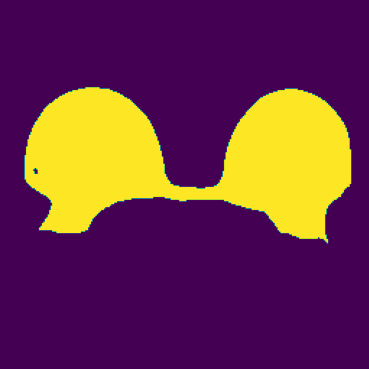}

 \includegraphics[width=0.15\textwidth]{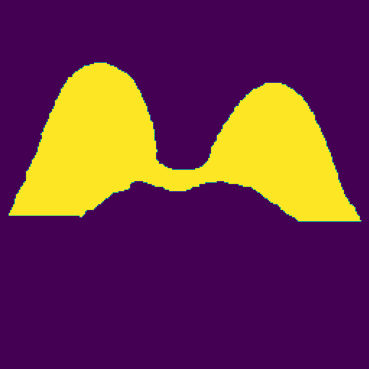} 
    \includegraphics[width=0.15\textwidth]{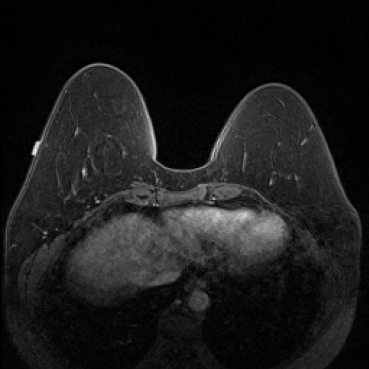} 
    \includegraphics[width=0.15\textwidth]{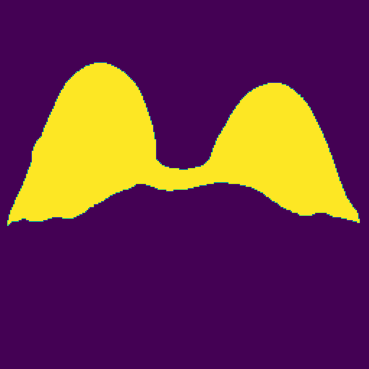}
     \includegraphics[width=0.15\textwidth]{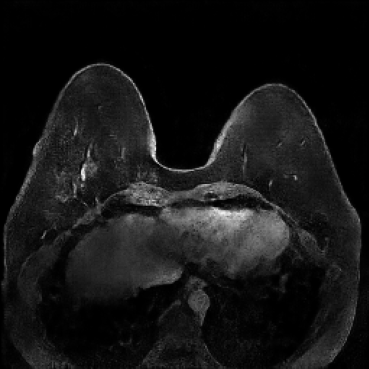} 
    \includegraphics[width=0.15\textwidth]{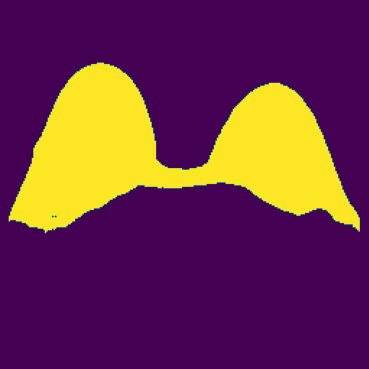}
    
    \caption{Comparation of segmentation results. The first column is the Ground Truth of semantic maps, the second column is the real images from test set, the third column is the segmentation results of the real images from test set, the fourth column is the synthetic images generated by Med-LSDM using semantic map from test set, and the fifth column is the segmentation results of the synthetic images.}
    \label{segmentation results}
\end{figure}

The goal of our research is not to generate fancy or visually appealing images, but rather to produce images that are useful for downstream tasks. Our focus is on creating synthetic data that have small domain gap with real data, helping to address the bottleneck of data scarcity in the field of machine learning for medical imaging. By generating images that are meaningful for specific tasks, we aim to improve the performance of models in clinical applications, ultimately driving advancements in medical image analysis where labeled data are often limited. In this experiment, We first train a 3D segmentation network using the images \( x \) and the corresponding semantic maps \( m \) from training set of the real Duke Breast dataset. The model and training details can be found in the Appendix \ref{tab:appendixB}. The purpose is to allow the network to learn how to segment anatomical structures from real MRT data. After completing the training phase, we evaluated the segmentation network on four different datasets as shown in Table~\ref{Test results from different data resources}: (1) real images from the Duke Breast dataset training set, (2) real images from the Duke Breast dataset test set, (3) synthetic images generated by our SegGuidedDiff model~\cite{konz2024anatomicallycontrollablemedicalimagegeneration} using semantic maps from the Duke Breast test set, and (4) synthetic images produced by our Med-LSDM model based on the same semantic maps. By evaluating the segmentation results on different real and synthetic images, we aim to investigate the effectiveness of the generative model in producing realistic images that could yield similar segmentation results to real data, suggesting that real and synthetic images lie in similar regions of the data distribution.

Table~\ref{Test results from different data resources} presents the segmentation performance, measured by the Dice similarity coefficient averaged over all semantic classes, across different data sources. The real training data achieve the highest Dice score of 0.750, which indicates that the segmentation model achieves a 75\% overlap between the predicted region and the ground truth. When tested on the real test set, the score falls slightly to 0.715, indicating good generalization.

In contrast, the segmentation result of synthetic data generated by SegGuidedDiff using 2D semantic map slices shows significantly lower performance (Dice = 0.602). Notably, SegGuidedDiff generates images slice-by-slice in 2D without modeling the full 3D spatial context. As a result, the synthesized volumes often suffer from inter-slice inconsistencies and unrealistic spatial structures, which negatively impact the downstream segmentation performance.

Our Med-LSDM-generated data, in contrast, achieves a Dice score of 0.70964, which is much higher than SegGuidedDiff and closely approaches the real data test performance. This indicates that our method's 3D generation strategy successfully captures anatomical coherence across slices and produces realistic volumetric data. It is also important to note that the synthetic volumes are directly saved in NIfTI format without any post-processing, demonstrating the practical applicability and robustness of our framework.

\begin{table}[h!]
\centering
\caption{Comparison of Dice scores achieved by the same segmentation network on different real and synthetic MR datasets}
\begin{tabular}{c|c}
 \toprule
   Data resource       & Dice \\ 
   \midrule
   Real MR data: Duke Breast training set          & 0.750 \\ 
   Real MR data: Duke Breast test set              & 0.715 \\ 
   Synthetic MR data: Generated by SegGuidedDiff  \cite{konz2024anatomicallycontrollablemedicalimagegeneration} & 0.602 \\ 
   Synthetic MR data: Generated by our approach (\textbf{ours})           & \textbf{0.710} \\ 
   \bottomrule
\end{tabular}

\label{Test results from different data resources}
\end{table}



The hypothesis of this experiment is that if the synthetic images generated by our model resemble real medical images in distribution, a segmentation network trained on real data should yield similar performance on synthetic data. By comparing the segmentation results across real and synthetic images, we can quantitatively assess the structural fidelity and semantic consistency of the generated data. The results confirm that segmentation performance on synthetic images is highly comparable to that on real images, indicating that our model successfully captures anatomically relevant structures. This suggests that synthetic data are not only realistic but also potentially beneficial as an augmentation strategy to improve segmentation tasks. Visualizations of the segmentation output are presented in Figure~\ref{segmentation results}.

\subsection{VQ-GAN Can Effectively Compress Images in a Perceptual Manner }

Our method builds upon the assumption that the latent space learned by the VQ-GAN can effectively compress spatial information while preserving its correspondence with the pixel space. This property ensures that conducting semantic diffusion within the latent space remains meaningful and consistent with the original image structure.

\begin{figure}[h]
\centering
 \begin{minipage}{0.15\textwidth}
        \centering
        \includegraphics[width=\textwidth]{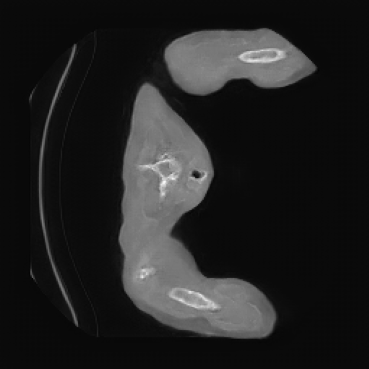}
    \end{minipage}%
    \begin{minipage}{0.15\textwidth}
        \centering
        \includegraphics[width=\textwidth]{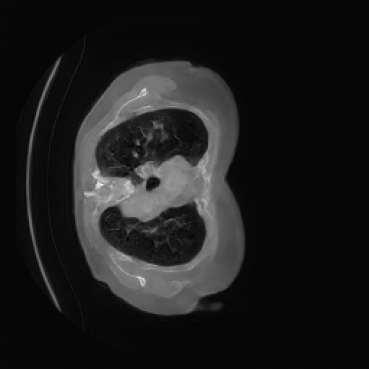} 
    \end{minipage}%
    \begin{minipage}{0.15\textwidth}
        \centering
        \includegraphics[width=\textwidth]{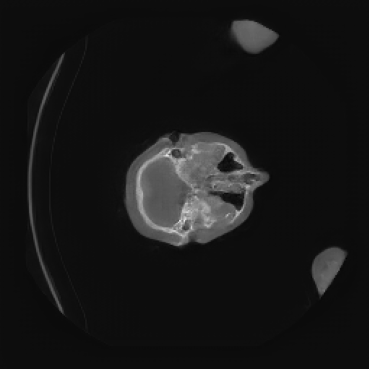}
    \end{minipage}%
      \begin{minipage}{0.15\textwidth}
        \centering
        \includegraphics[width=\textwidth]{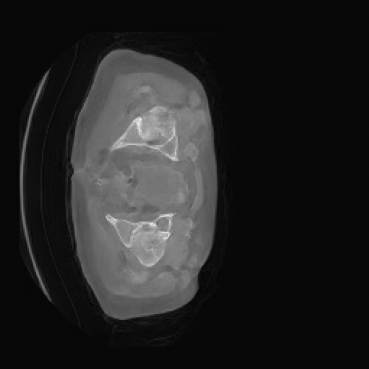}
    \end{minipage}%
      \begin{minipage}{0.15\textwidth}
        \centering
         \includegraphics[width=\textwidth]{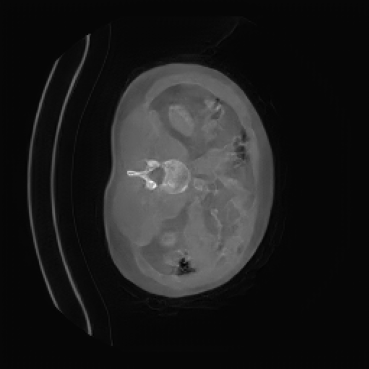}
    \end{minipage}%

    \centering
         \begin{minipage}{0.15\textwidth}
        \centering
        \includegraphics[width=\textwidth]{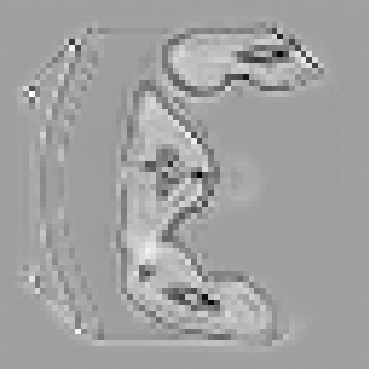}
    \end{minipage}%
    \begin{minipage}{0.15\textwidth}
        \centering
        \includegraphics[width=\textwidth]{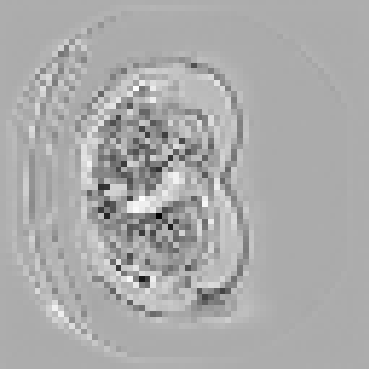} 
    \end{minipage}%
    \begin{minipage}{0.15\textwidth}
        \centering
        \includegraphics[width=\textwidth]{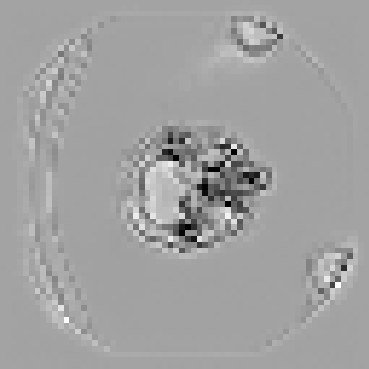}
    \end{minipage}%
      \begin{minipage}{0.15\textwidth}
        \centering
        \includegraphics[width=\textwidth]{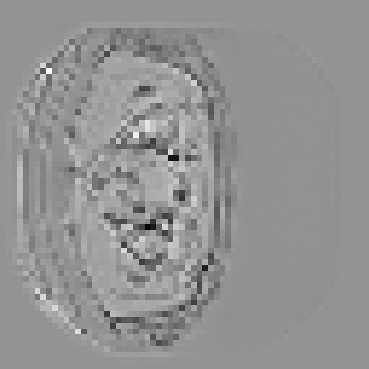}
    \end{minipage}%
      \begin{minipage}{0.15\textwidth}
        \centering
         \includegraphics[width=\textwidth]{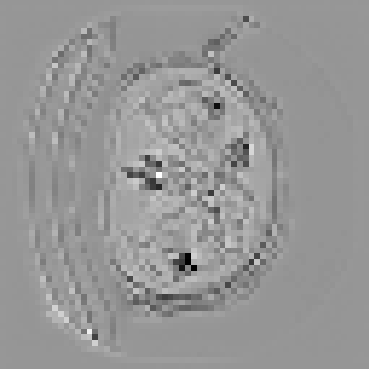}
    \end{minipage}%
    
    \caption{The corresponding 2D slices of reconstructed images $\hat{{x}}$ (256*256) in the first row and 2D slices of latent representation $\hat{{z}}$ (64*64) in the second row}
    \label{latent space}
\end{figure}

In our approach, we employ a compression factor of 4 to encode 3D medical images into the latent space. This level of compression reduces the spatial resolution of the original image (e.g., from 256×256 to 64×64 per slice), while still preserving the essential anatomical structures and semantic content. The latent representation at this scale offers a favorable trade-off between dimensionality reduction and semantic fidelity. Specifically, although fine-grained textures are simplified, key structural patterns (e.g., organ boundaries, lesion shapes) remain visually distinguishable and semantically coherent. As the compression rate increases, the latent representations become progressively more abstract. With lower compression (e.g., 2× or 4×), the latent features preserve key anatomical structures and spatial layouts, making them beneficial for our semantic image synthesis task in latent space, as the model can operate on compressed representations that retain sufficient semantic information without being overwhelmed by high-frequency noise. In contrast, higher compression rates (e.g., 8× or above) lead to a loss of fine-grained details and reduced semantic fidelity. The choice of a 4× compression thus ensures that the latent features are compact and meaningful, facilitating effective conditional generation while significantly reducing computational overhead. We therefore adopt a 4× compression as a compromise between computational efficiency and semantic preservation. The corresponding 2D slices in image space and latent space are shown in Figure \ref{latent space}.

Most bits in a digital medical image represent imperceptible or semantically irrelevant details~\cite{ho2020denoisingdiffusionprobabilisticmodels}. To address this, we employ a VQ-GAN as a perceptual compressor that transforms high-resolution medical images into a compact latent space while preserving essential semantic structures. This compressed representation significantly reduces spatial dimensionality, thereby lowering the computational and memory complexity of subsequent generative modeling. Additionally, operating in the perceptually meaningful latent space enables the latent diffusion model to focus on high-level semantic generation, leading to improved synthesis quality. Finally, the decoder reconstructs high-resolution images from the latent space. By separating the compression and generation stages, our framework enables both computationally efficient and semantically faithful 3D medical image synthesis.

\subsection{Med-LSDM outperform 2D SegGuidedDiff Model }

 \begin{figure}[htb]
	\centering
	\includegraphics[scale=.45]{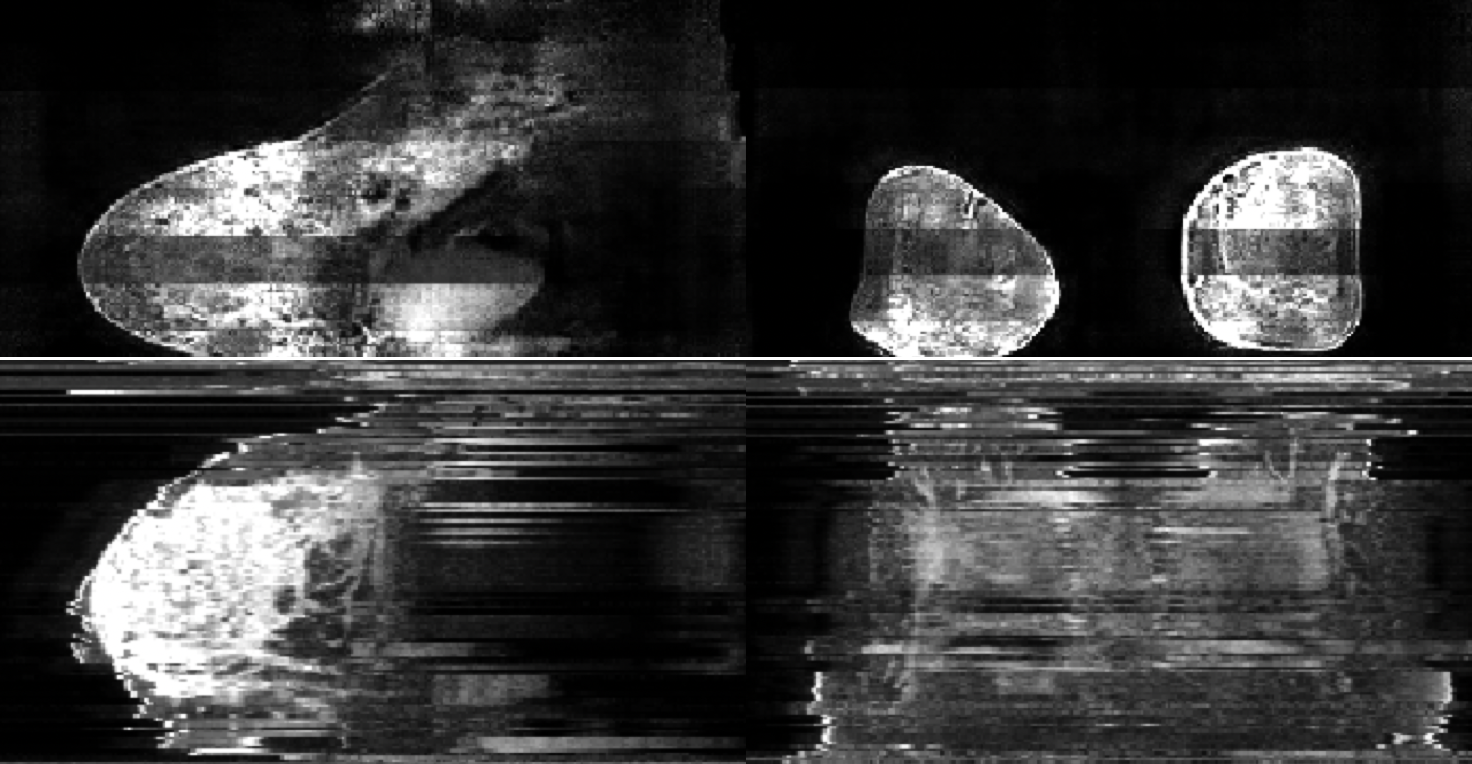}
	\caption{The examples of slices from coronal and sagittal planes from our model (top row) and SegGuidedDiff Model (bottom row)}
	\label{coronal and sagittal plane}
\end{figure} 


We conduct a comprehensive evaluation of our proposed Med-LSDM model against the SegGuidedDiff baseline across three public datasets: AutoPET, SynthRAD2023, and Duke Breast. To do this, we compute multiple quantitative metrics that measure image fidelity and structural similarity between Real images from the original datasets, and Synthetic images generated from the corresponding semantic maps using each generative model.
Table~\ref{tab:comparison} presents a detailed quantitative comparison, while Figure~\ref{coronal and sagittal plane} illustrates representative coronal and sagittal views of generated 3D volumes.

\begin{table}[!ht]
\centering
\caption{Quantitative comparison between Med-LSDM (ours) and SegGuidedDiff across three datasets.}
\label{tab:comparison}
\renewcommand{\arraystretch}{1.2}
\begin{tabular}{ll|cccccc}
\toprule
\textbf{Method} & \textbf{Dataset} & \textbf{FID} ↓ & \textbf{3D-FID} ↓ & \textbf{LPIPS} ↓ & \textbf{SSIM} ↑ & \textbf{RMSE} ↓ & \textbf{PSNR} ↑ \\
\midrule
SegGuidedDiff   & AutoPET   & 197.157 & 0.0235 & 0.19 & 0.9300 & 0.19 & 15.14 \\
SegGuidedDiff   & SynthRAD2023  & 196.780 & 0.0362 & 0.19 & 0.9800 & 0.28 & 11.65 \\
SegGuidedDiff   & Duke Breast    & 197.060 & 0.0145 & 0.24 & 0.7090 & 0.41 & 7.74 \\
\midrule
Med-LSDM (ours) & AutoPET   & 196.780 & 0.0090 & 0.19 & 0.8374 & 0.13 & 17.75 \\
Med-LSDM (ours) & SynthRAD2023  & 198.030 & 0.0061 & 0.27 & 0.6147 & 0.36 & 8.89 \\
Med-LSDM (ours) & Duke Breast & 193.620 & 0.0054 & 0.29 & 0.9390 & 0.21 & 14.87 \\
\bottomrule
\end{tabular}
\end{table}

As shown in Table~\ref{tab:comparison}, the performance comparison demonstrates dataset-dependent effectiveness with notably different patterns across the three evaluation scenarios.
On AutoPET Dataset, Med-LSDM achieves superior performance across most metrics, particularly excelling in 3D volumetric consistency with a substantial improvement in 3D-FID (0.0090 vs. 0.0235). The model also demonstrates better pixel-wise accuracy with lower RMSE (0.13 vs. 0.19) and higher PSNR (17.75 vs. 15.14), while maintaining comparable perceptual similarity (LPIPS) and distribution alignment (FID).

On Duke Breast Dataset, Our method shows significant improvements across all key metrics, achieving notably lower FID (193.620 vs. 197.060) and 3D-FID (0.0054 vs. 0.0145). The substantial improvement in SSIM (0.9390 vs. 0.7090) and PSNR (14.87 vs. 7.74) demonstrates the effectiveness of Med-LSDM for breast MRI synthesis.

However, Med-LSDM exhibits suboptimal performance compared to SegGuidedDiff on SynthRAD2023 Dataset across most traditional metrics, with notably degraded SSIM (0.6147 vs. 0.9800), higher LPIPS (0.27 vs. 0.19), and lower PSNR (8.89 vs. 11.65). The only improvement observed is in 3D-FID (0.0061 vs. 0.0362), suggesting better volumetric consistency despite the compromised slice-wise quality. The contrast in performance on SynthRAD2023 reveals fundamental challenges in cross-modal medical image synthesis. Unlike AutoPET and Duke Breast, which involve intra-modal generation tasks, SynthRAD2023 requires generating MR images from CT-derived semantic maps, where is a significantly more challenging scenario due to several critical factors. First, cross-modal semantic inconsistency presents a fundamental limitation. CT and MR modalities exhibit dramatically different tissue contrasts and anatomical visibility characteristics.
Second, the model complexity versus data availability trade-off becomes particularly pronounced in the SynthRAD2023 scenario. Our Med-LSDM architecture contains significantly more parameters than the simpler SegGuidedDiff baseline. Although this increased capacity enables superior performance on datasets with adequate training samples (AutoPET and Duke Breast), it becomes a liability when training data is limited, as in SynthRAD2023. The models especially the VQ-GAN overfit to the limited cross-modal training examples.



Figure~\ref{coronal and sagittal plane} further supports these findings: the images generated by Med-LSDM preserve spatial coherence across coronal and sagittal planes, whereas those from SegGuidedDiff show noticeable inconsistencies. These artifacts likely result from the 2D nature of SegGuidedDiff, which fails to account for 3D spatial relationships. In contrast, our model’s volumetric diffusion framework inherently preserves anatomical consistency, highlighting its advanta in synthesizing high-fidelity 3D medical images.


\section{Discussion and Conclusion}

In this study, we propose Med-LSDM (Latent Semantic Diffusion Model) framework for 3D semantic image synthesis for medical imaging. The architecture of the network consists of two key components: a 3D VQ-GAN and a 3D Semantic Diffusion Model. We implement the diffusion model in the learned latent space of VQ-GAN. Our aim is to generate synthetic medical data from semantic maps as a data augmentation method. 

The effect of our framework was validated on three challenging datasets: AutoPET, SynthRAD2023 and Duke Breast dataset. The proposed framework achieves promising results for intra-modal synthesis tasks with sufficient training data, where the model's capacity for learning complex anatomical relationships translates to superior 3D consistency and image quality. However, for cross-modal synthesis scenarios or limited data regimes, Our model still has a lot of room for improvement.
The superior 3D-FID scores across all datasets highlight a key strength of our volumetric approach: even when slice-wise metrics are compromised, the preservation of 3D spatial relationships remains robust. This characteristic may be particularly valuable for clinical applications where anatomical consistency across viewing planes is critical, though the trade-offs with local image fidelity must be carefully considered based on specific use cases.

Our experiments prove that the model could effectively bridge the domain gap between real and generated data, as evidenced by the high similarity in segmentation results on both real and synthetic images. This finding highlights the potential of using synthetic data to augment real datasets in medical image analysis tasks. To the best of our knowledge, it is the first work to implement semantic medical image synthesis in latent space. By removing personally identifiable information and concentrating on medically pertinent, anonymized data, this method adheres to legal and ethical guidelines while also reducing the risks to privacy that come with data exposure during the processes of sharing or publishing.

Our model is currently limited to generating 32-slice images due to computational constraints. In real-world medical applications, imaging studies often consist of hundreds of slices that cover more comprehensive anatomical regions. 
Addressing this limitation will be a key focus of future work, with the goal of generating full 3D medical images from end-to-end, using complete semantic maps as input. Furthermore, expanding the model’s capability to work with multi-modal data, such as MRI, CT, and ultrasound, will be essential for achieving broader generalization and enhancing the flexibility of the system. By generating images across various modalities from input labels, the model could help address the challenges of multi-modality imaging in clinical practice. Modality-aware semantic mapping could address cross-modal synthesis challenges by developing conditioning representations that better capture target modality characteristics. Progressive training strategies might leverage large intra-modal datasets for pre-training before fine-tuning on limited cross-modal data. Additionally, regularization techniques specifically designed for small dataset scenarios could help large models generalize more effectively in data-limited regimes.

\bibliographystyle{unsrt}  
\bibliography{references}  

\begin{thebibliography}{10}

\bibitem{MedicalImageAnalysis}
Balasubramaniam .S, Prasanth Ap, Satheesh Kumar, and V.~Kavitha.
\newblock Medical image analysis based on deep learning approach for early diagnosis of diseases, 03 2024.

\bibitem{Litjens_2017}
Geert Litjens, Thijs Kooi, Babak~Ehteshami Bejnordi, Arnaud Arindra~Adiyoso Setio, Francesco Ciompi, Mohsen Ghafoorian, Jeroen~A.W.M. van~der Laak, Bram van Ginneken, and Clara~I. Sánchez.
\newblock A survey on deep learning in medical image analysis.
\newblock {\em Medical Image Analysis}, 42:60–88, December 2017.

\bibitem{he2022transformersmedicalimageanalysis}
Kelei He, Chen Gan, Zhuoyuan Li, Islem Rekik, Zihao Yin, Wen Ji, Yang Gao, Qian Wang, Junfeng Zhang, and Dinggang Shen.
\newblock Transformers in medical image analysis: A review, 2022.

\bibitem{moeskops2022deep}
Pim Moeskops, Max~A Viergever, Juan~M I{\~n}esta, and Josien~P Pluim.
\newblock Deep learning techniques for medical image segmentation: Achievements and challenges.
\newblock {\em Journal of Imaging Informatics in Medicine}, 10:123--139, 2022.

\bibitem{luca2022impact}
Andreea~Roxana Luca, Tudor~Florin Ursuleanu, and Liliana Gheorghe.
\newblock Impact of quality, type and volume of data used by deep learning models in the analysis of medical images.
\newblock {\em Informatics in Medicine Unlocked}, 29:100911, 2022.

\bibitem{prevedello2019challenges}
Luciano~M Prevedello, Safwan~S Halabi, George Shih, Carol~C Wu, Marc~D Kohli, Falgun~H Chokshi, Bradley~J Erickson, Jayashree Kalpathy-Cramer, Katherine~P Andriole, and Adam~E Flanders.
\newblock Challenges related to artificial intelligence research in medical imaging and the importance of image analysis competitions.
\newblock {\em Radiology: Artificial Intelligence}, 1(1):e180031, 2019.

\bibitem{ho2020denoisingdiffusionprobabilisticmodels}
Jonathan Ho, Ajay Jain, and Pieter Abbeel.
\newblock Denoising diffusion probabilistic models, 2020.

\bibitem{zhang2021shifting}
Angela Zhang, Lei Xing, James Zou, and Joseph~C Wu.
\newblock Shifting machine learning for healthcare from development to deployment and from models to data.
\newblock {\em Nature}, 580(252-256), 2021.

\bibitem{chen2021gan_review}
X.~Chen, Y.~Zhang, Y.~Li, and C.~Sun.
\newblock The use of generative adversarial networks in medical image augmentation: A review.
\newblock {\em Neural Computing and Applications}, 33:15227--15246, 2021.

\bibitem{hippa}
National~Archives Office of~the Federal~Register and Records Administration.
\newblock Health insurance portability and accountability act of 1996, August 21, 1996.

\bibitem{Ma_2024}
Jun Ma, Yuting He, Feifei Li, Lin Han, Chenyu You, and Bo~Wang.
\newblock Segment anything in medical images.
\newblock {\em Nature Communications}, 15(1), 2024.

\bibitem{isola2018imagetoimagetranslationconditionaladversarial}
Phillip Isola, Jun-Yan Zhu, Tinghui Zhou, and Alexei~A. Efros.
\newblock Image-to-image translation with conditional adversarial networks, 2018.

\bibitem{wang2018pix2pixHD}
Ting-Chun Wang, Ming-Yu Liu, Jun-Yan Zhu, Andrew Tao, Jan Kautz, and Bryan Catanzaro.
\newblock High-resolution image synthesis and semantic manipulation with conditional gans.
\newblock In {\em Proceedings of the IEEE Conference on Computer Vision and Pattern Recognition}, 2018.

\bibitem{esser2021tamingtransformershighresolutionimage}
Patrick Esser, Robin Rombach, and Björn Ommer.
\newblock Taming transformers for high-resolution image synthesis, 2021.

\bibitem{Armanious_2020}
Karim Armanious, Chenming Jiang, Marc Fischer, Thomas Küstner, Tobias Hepp, Konstantin Nikolaou, Sergios Gatidis, and Bin Yang.
\newblock Medgan: Medical image translation using gans.
\newblock {\em Computerized Medical Imaging and Graphics}, 79:101684, January 2020.

\bibitem{su2023dualdiffusionimplicitbridges}
Xuan Su, Jiaming Song, Chenlin Meng, and Stefano Ermon.
\newblock Dual diffusion implicit bridges for image-to-image translation, 2023.

\bibitem{DBLP:journals/corr/abs-1903-07291}
Taesung Park, Ming{-}Yu Liu, Ting{-}Chun Wang, and Jun{-}Yan Zhu.
\newblock Semantic image synthesis with spatially-adaptive normalization.
\newblock {\em CoRR}, abs/1903.07291, 2019.

\bibitem{Zhu_2020}
Peihao Zhu, Rameen Abdal, Yipeng Qin, and Peter Wonka.
\newblock Sean: Image synthesis with semantic region-adaptive normalization.
\newblock In {\em 2020 IEEE/CVF Conference on Computer Vision and Pattern Recognition (CVPR)}. IEEE, June 2020.

\bibitem{karras2020analyzingimprovingimagequality}
Tero Karras, Samuli Laine, Miika Aittala, Janne Hellsten, Jaakko Lehtinen, and Timo Aila.
\newblock Analyzing and improving the image quality of stylegan, 2020.

\bibitem{eskandar2021usisunsupervisedsemanticimage}
George Eskandar, Mohamed Abdelsamad, Karim Armanious, and Bin Yang.
\newblock Usis: Unsupervised semantic image synthesis, 2021.

\bibitem{wang2022semanticimagesynthesisdiffusion}
Weilun Wang, Jianmin Bao, Wengang Zhou, Dongdong Chen, Dong Chen, Lu~Yuan, and Houqiang Li.
\newblock Semantic image synthesis via diffusion models, 2022.

\bibitem{rombach2022highresolutionimagesynthesislatent}
Robin Rombach, Andreas Blattmann, Dominik Lorenz, Patrick Esser, and Björn Ommer.
\newblock High-resolution image synthesis with latent diffusion models, 2022.

\bibitem{goodfellow2014generative}
Ian Goodfellow, Jean Pouget-Abadie, Mehdi Mirza, Bing Xu, David Warde-Farley, Sherjil Ozair, Aaron Courville, and Yoshua Bengio.
\newblock Generative adversarial nets.
\newblock {\em Advances in neural information processing systems}, 27, 2014.

\bibitem{Yi_2019}
Xin Yi, Ekta Walia, and Paul Babyn.
\newblock Generative adversarial network in medical imaging: A review.
\newblock {\em Medical Image Analysis}, 58:101552, December 2019.

\bibitem{qasim2021redganattackingclassimbalance}
Ahmad~B Qasim, Ivan Ezhov, Suprosanna Shit, Oliver Schoppe, Johannes~C Paetzold, Anjany Sekuboyina, Florian Kofler, Jana Lipkova, Hongwei Li, and Bjoern Menze.
\newblock Red-gan: Attacking class imbalance via conditioned generation. yet another perspective on medical image synthesis for skin lesion dermoscopy and brain tumor mri, 2021.

\bibitem{konz2024anatomicallycontrollablemedicalimagegeneration}
Nicholas Konz, Yuwen Chen, Haoyu Dong, and Maciej~A. Mazurowski.
\newblock Anatomically-controllable medical image generation with segmentation-guided diffusion models, 2024.

\bibitem{khader2023medicaldiffusiondenoisingdiffusion}
Firas Khader, Gustav Mueller-Franzes, Soroosh~Tayebi Arasteh, Tianyu Han, Christoph Haarburger, Maximilian Schulze-Hagen, Philipp Schad, Sandy Engelhardt, Bettina Baessler, Sebastian Foersch, Johannes Stegmaier, Christiane Kuhl, Sven Nebelung, Jakob~Nikolas Kather, and Daniel Truhn.
\newblock Medical diffusion: Denoising diffusion probabilistic models for 3d medical image generation, 2023.

\bibitem{Dorjsembe_2024}
Zolnamar Dorjsembe, Hsing-Kuo Pao, Sodtavilan Odonchimed, and Furen Xiao.
\newblock Conditional diffusion models for semantic 3d brain mri synthesis.
\newblock {\em IEEE Journal of Biomedical and Health Informatics}, 28(7):4084–4093, July 2024.

\bibitem{ramachandran2017searchingactivationfunctions}
Prajit Ramachandran, Barret Zoph, and Quoc~V. Le.
\newblock Searching for activation functions, 2017.

\bibitem{nair2010rectified}
Vinod Nair and Geoffrey~E Hinton.
\newblock Rectified linear units improve restricted boltzmann machines.
\newblock In {\em Proceedings of the 27th international conference on machine learning (ICML-10)}, pages 807--814, 2010.

\bibitem{gatidis2022whole}
Sergios Gatidis, Tobias Hepp, Marcel Fr{\"u}h, Christian La~Foug{\`e}re, Konstantin Nikolaou, Christina Pfannenberg, Bernhard Sch{\"o}lkopf, Thomas K{\"u}stner, Clemens Cyran, and Daniel Rubin.
\newblock A whole-body fdg-pet/ct dataset with manually annotated tumor lesions.
\newblock {\em Scientific Data}, 9(1):601, 2022.

\bibitem{Thummerer_2023}
Adrian Thummerer, Erik van~der Bijl, Arthur Galapon, Joost J.~C. Verhoeff, Johannes~A. Langendijk, Stefan Both, Cornelis (Nico) A.~T. van~den Berg, and Matteo Maspero.
\newblock Synthrad2023 grand challenge dataset: Generating synthetic ct for radiotherapy.
\newblock {\em Medical Physics}, 50(7):4664–4674, June 2023.

\bibitem{breastcancer}
A.~Saha, M.R. Harowicz, and L.J. Grimm.
\newblock A machine learning approach to radiogenomics of breast cancer: a study of 922 subjects and 529 dce-mri features.
\newblock {\em Journal of Magnetic Resonance Imaging}, 49:508–516, 2018.

\bibitem{Wasserthal_2023}
Jakob Wasserthal, Hanns-Christian Breit, Manfred~T. Meyer, Maurice Pradella, Daniel Hinck, Alexander~W. Sauter, Tobias Heye, Daniel~T. Boll, Joshy Cyriac, Shan Yang, Michael Bach, and Martin Segeroth.
\newblock Totalsegmentator: Robust segmentation of 104 anatomic structures in ct images.
\newblock {\em Radiology: Artificial Intelligence}, 5(5), September 2023.

\bibitem{9770375}
Li~Sun, Junxiang Chen, Yanwu Xu, Mingming Gong, Ke~Yu, and Kayhan Batmanghelich.
\newblock Hierarchical amortized gan for 3d high resolution medical image synthesis.
\newblock {\em IEEE Journal of Biomedical and Health Informatics}, 26(8):3966--3975, 2022.

\bibitem{chen2019med3dtransferlearning3d}
Sihong Chen, Kai Ma, and Yefeng Zheng.
\newblock Med3d: Transfer learning for 3d medical image analysis, 2019.

\end{thebibliography}


\newpage

\appendix
\section*{Appendix}\label{Appendix}
\addcontentsline{toc}{section}{Appendix}

\section{Number of model parameters of our architecture}

\begin{table}[h]
    \centering
    \caption{Trainable Parameters of VQ-GAN Components}
    \label{modelparameter}
    \begin{tabular}{cc|cccccc}
        \toprule
         & Name & Encoder & Decoder & 2D-Dis& 3D-Dis& Perceptual-model (VGG) \\
        \midrule
            & \#param & 441.5K & 948.6K & 2.8M & 11.0M & 14.7M\\

        \bottomrule
    \end{tabular}
\end{table}

\section{Detailed information of Segmentation Network}\label{tab:appendixB}

\begin{table}[ht]
\centering
\caption{U-Net3D Model Parameters and Training Configuration}
\label{tab:unet3d_params_config}
\begin{tabular}{|c|c|}
\hline
\textbf{Parameter} & \textbf{Value} \\ \hline
\multicolumn{2}{|c|}{\textbf{Model Parameters}} \\ \hline
Number of Classes & 2 \\ \hline
Input Channels & 1 \\ \hline
Kernel Size & 3 \\ \hline
Normalization & InstanceNorm3d \\ \hline
Optimizer & Adam \\ \hline
Learning Rate & 0.00005 \\ \hline
Loss Functions & Dice Coefficient Loss and Cross Entropy Loss \\ \hline
\multicolumn{2}{|c|}{\textbf{Training Configuration}} \\ \hline
Batch Size & 4 \\ \hline
Patch Size & (64, 64, 64) \\ \hline
Number of Epochs & 10 \\ \hline
Dataset & Duke Breast training set \\ \hline

\end{tabular}
\end{table}

\section{Conditional AutoPET Progressive Generation}

\begin{figure}[h]
    \centering
          \begin{minipage}{0.1\textwidth}
        \centering
        \includegraphics[width=\textwidth]{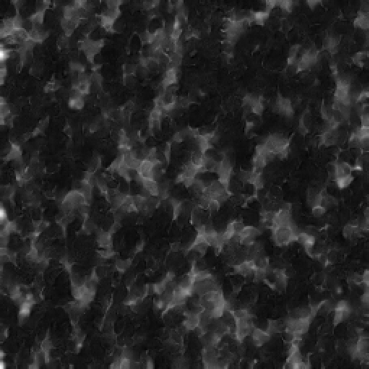}
    \end{minipage}%
    \begin{minipage}{0.1\textwidth}
        \centering
        \includegraphics[width=\textwidth]{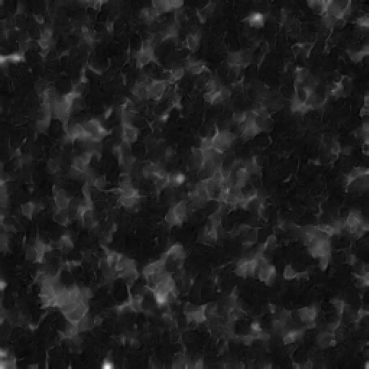} 
    \end{minipage}%
    \begin{minipage}{0.1\textwidth}
        \centering
        \includegraphics[width=\textwidth]{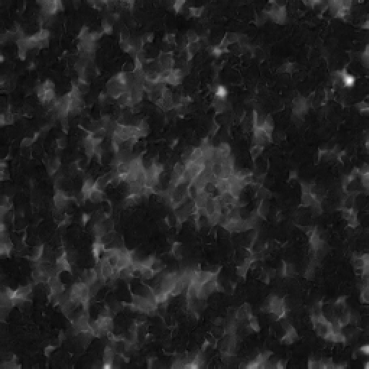}
    \end{minipage}%
      \begin{minipage}{0.1\textwidth}
        \centering
        \includegraphics[width=\textwidth]{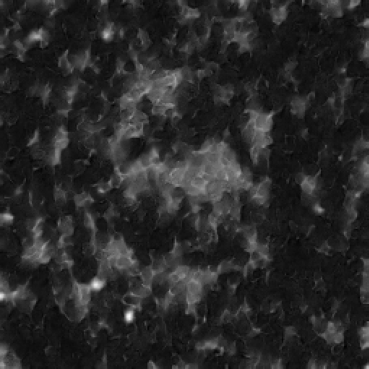}
    \end{minipage}%
      \begin{minipage}{0.1\textwidth}
        \centering
        \includegraphics[width=\textwidth]{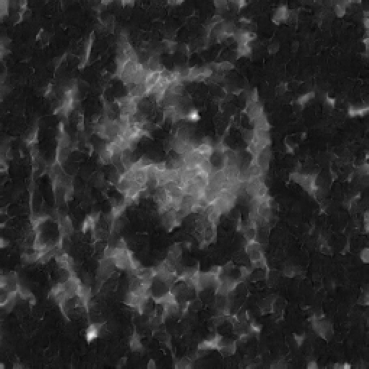}
    \end{minipage}%
      \begin{minipage}{0.1\textwidth}
        \centering
        \includegraphics[width=\textwidth]{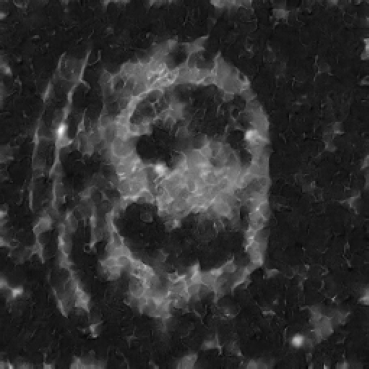}
    \end{minipage}%
      \begin{minipage}{0.1\textwidth}
        \centering
        \includegraphics[width=\textwidth]{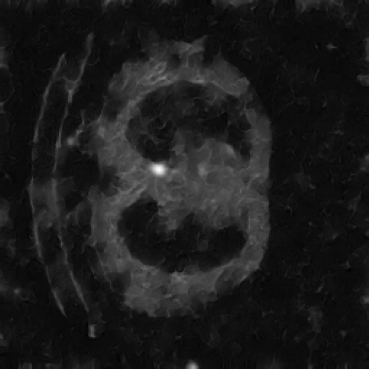}
    \end{minipage}%
      \begin{minipage}{0.1\textwidth}
        \centering
        \includegraphics[width=\textwidth]{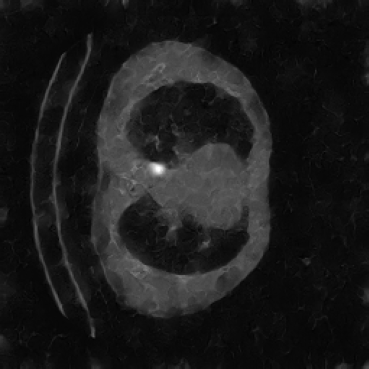}
    \end{minipage}%
      \begin{minipage}{0.1\textwidth}
        \centering
        \includegraphics[width=\textwidth]{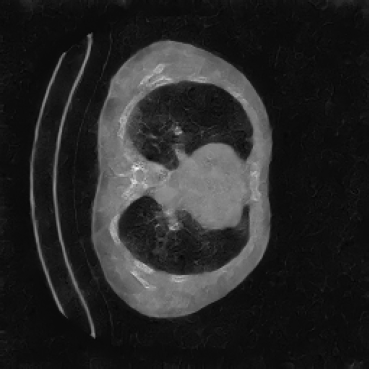}
    \end{minipage}%
      \begin{minipage}{0.1\textwidth}
        \centering
        \includegraphics[width=\textwidth]{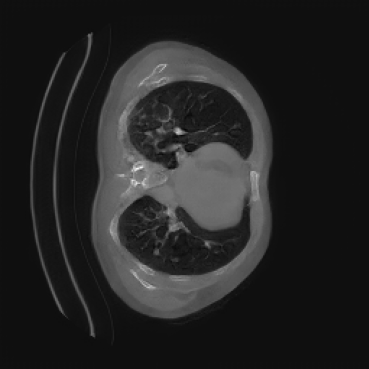}
    \end{minipage}%


            \begin{minipage}{0.1\textwidth}
        \centering
        \includegraphics[width=\textwidth]{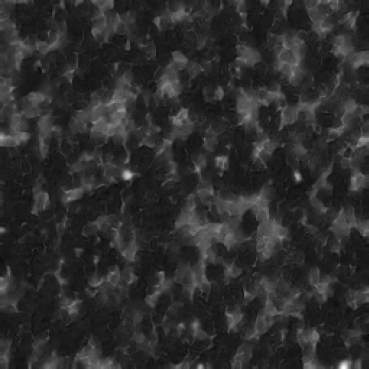}
    \end{minipage}%
    \begin{minipage}{0.1\textwidth}
        \centering
        \includegraphics[width=\textwidth]{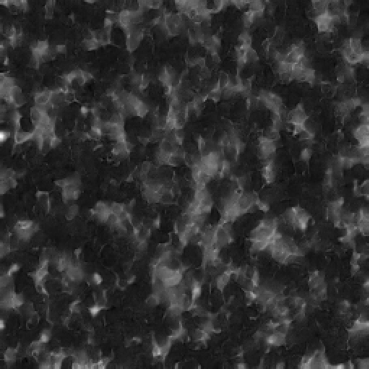} 
    \end{minipage}%
    \begin{minipage}{0.1\textwidth}
        \centering
        \includegraphics[width=\textwidth]{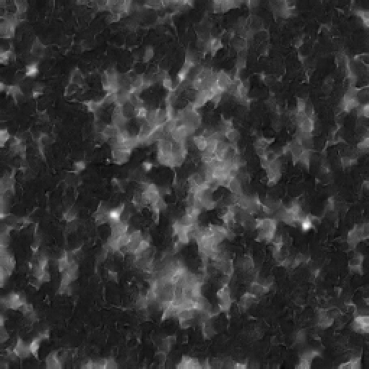}
    \end{minipage}%
      \begin{minipage}{0.1\textwidth}
        \centering
        \includegraphics[width=\textwidth]{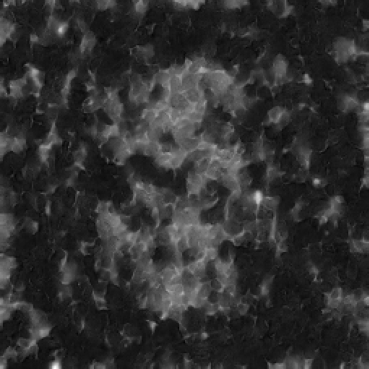}
    \end{minipage}%
      \begin{minipage}{0.1\textwidth}
        \centering
        \includegraphics[width=\textwidth]{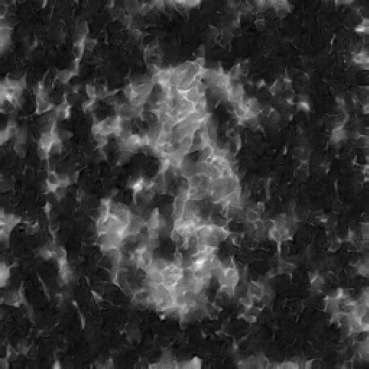}
    \end{minipage}%
      \begin{minipage}{0.1\textwidth}
        \centering
        \includegraphics[width=\textwidth]{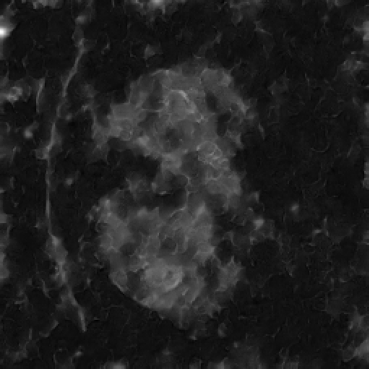}
    \end{minipage}%
      \begin{minipage}{0.1\textwidth}
        \centering
        \includegraphics[width=\textwidth]{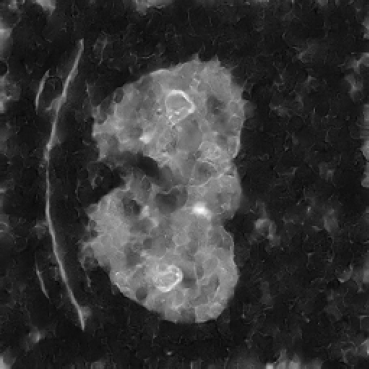}
    \end{minipage}%
      \begin{minipage}{0.1\textwidth}
        \centering
        \includegraphics[width=\textwidth]{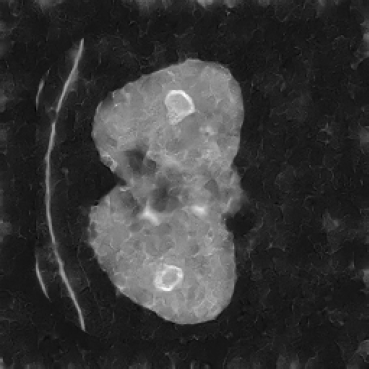}
    \end{minipage}%
      \begin{minipage}{0.1\textwidth}
        \centering
        \includegraphics[width=\textwidth]{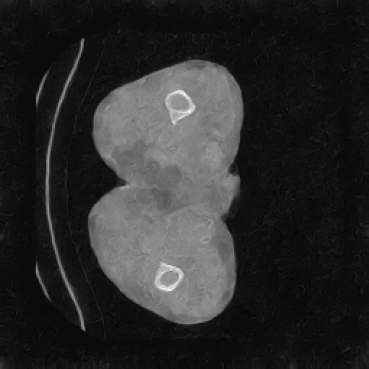}
    \end{minipage}%
      \begin{minipage}{0.1\textwidth}
        \centering
        \includegraphics[width=\textwidth]{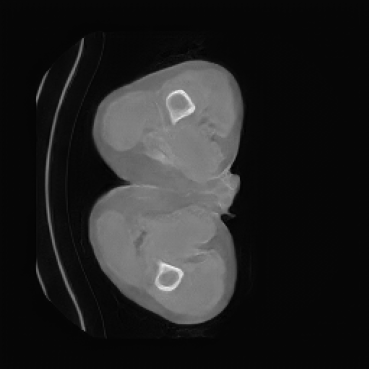}
    \end{minipage}%

            \begin{minipage}{0.1\textwidth}
        \centering
        \includegraphics[width=\textwidth]{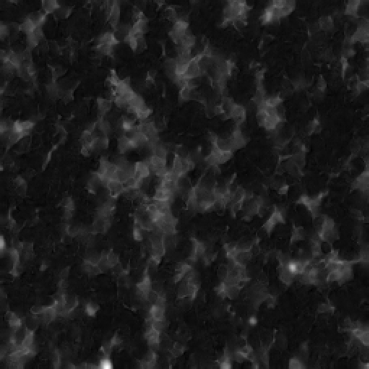}
    \end{minipage}%
    \begin{minipage}{0.1\textwidth}
        \centering
        \includegraphics[width=\textwidth]{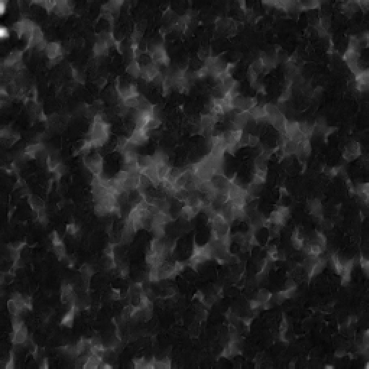} 
    \end{minipage}%
    \begin{minipage}{0.1\textwidth}
        \centering
        \includegraphics[width=\textwidth]{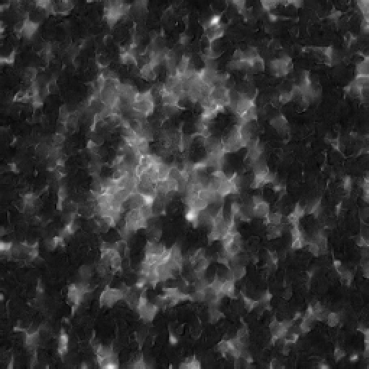}
    \end{minipage}%
      \begin{minipage}{0.1\textwidth}
        \centering
        \includegraphics[width=\textwidth]{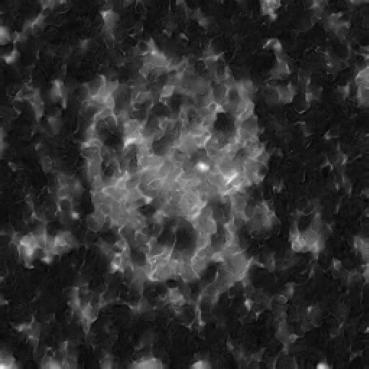}
    \end{minipage}%
      \begin{minipage}{0.1\textwidth}
        \centering
        \includegraphics[width=\textwidth]{ 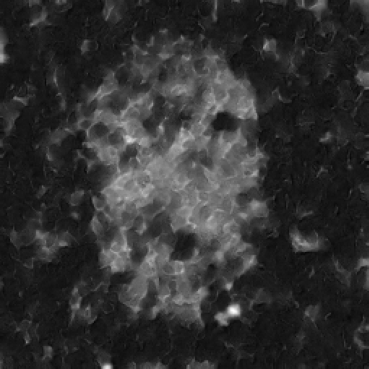}
    \end{minipage}%
      \begin{minipage}{0.1\textwidth}
        \centering
        \includegraphics[width=\textwidth]{ 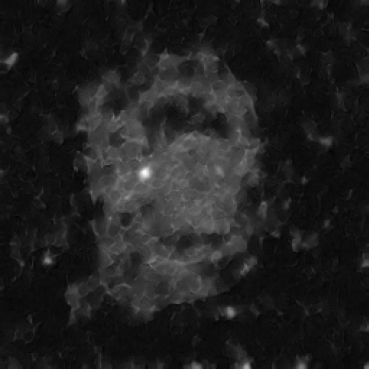}
    \end{minipage}%
      \begin{minipage}{0.1\textwidth}
        \centering
        \includegraphics[width=\textwidth]{ 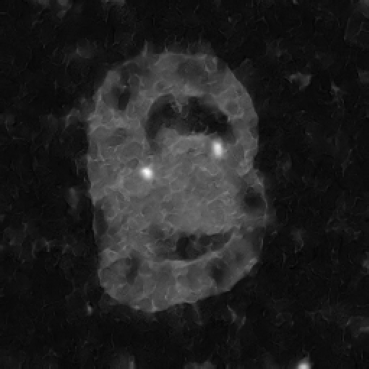}
    \end{minipage}%
      \begin{minipage}{0.1\textwidth}
        \centering
        \includegraphics[width=\textwidth]{ 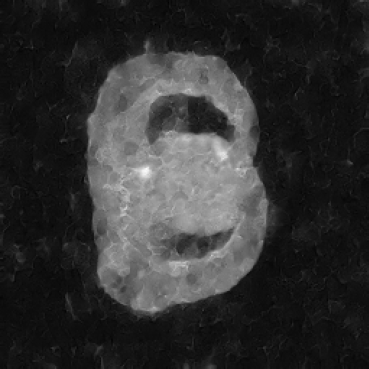}
    \end{minipage}%
      \begin{minipage}{0.1\textwidth}
        \centering
        \includegraphics[width=\textwidth]{ 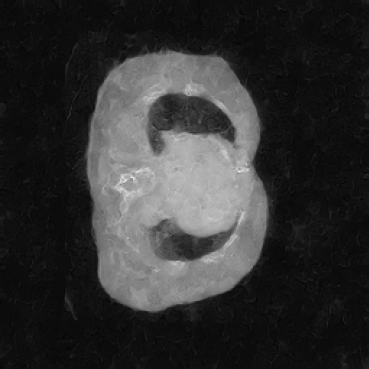}
    \end{minipage}%
      \begin{minipage}{0.1\textwidth}
        \centering
        \includegraphics[width=\textwidth]{ 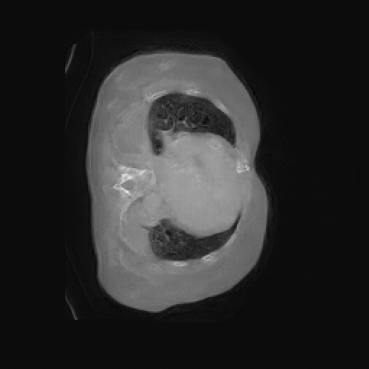}
    \end{minipage}%

            \begin{minipage}{0.1\textwidth}
        \centering
        \includegraphics[width=\textwidth]{ 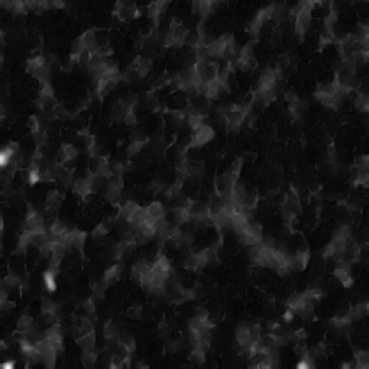}
    \end{minipage}%
    \begin{minipage}{0.1\textwidth}
        \centering
        \includegraphics[width=\textwidth]{ 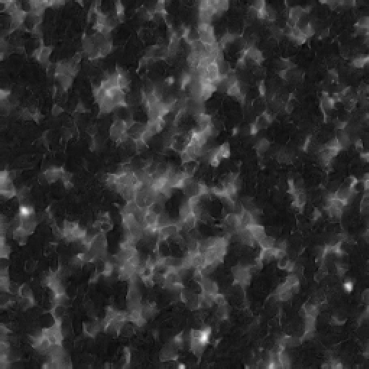} 
    \end{minipage}%
    \begin{minipage}{0.1\textwidth}
        \centering
        \includegraphics[width=\textwidth]{ 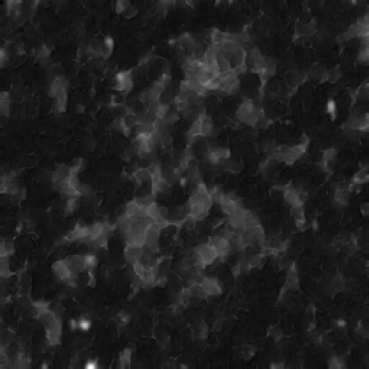}
    \end{minipage}%
      \begin{minipage}{0.1\textwidth}
        \centering
        \includegraphics[width=\textwidth]{ 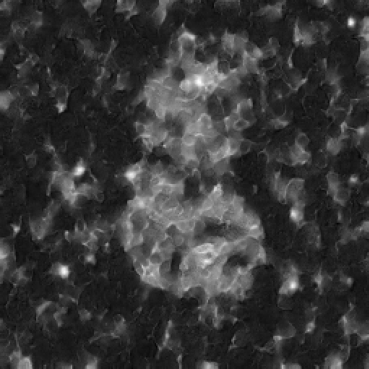}
    \end{minipage}%
      \begin{minipage}{0.1\textwidth}
        \centering
        \includegraphics[width=\textwidth]{ 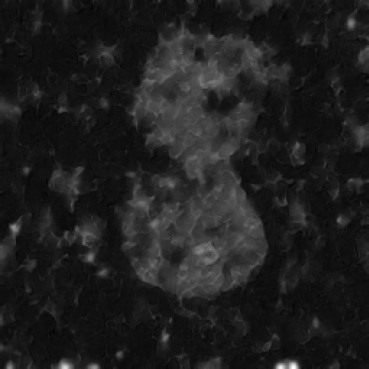}
    \end{minipage}%
      \begin{minipage}{0.1\textwidth}
        \centering
        \includegraphics[width=\textwidth]{ 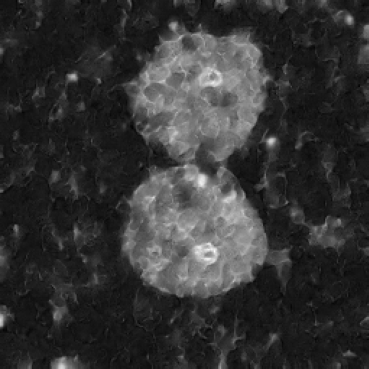}
    \end{minipage}%
      \begin{minipage}{0.1\textwidth}
        \centering
        \includegraphics[width=\textwidth]{ 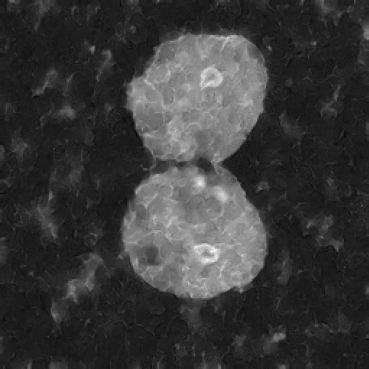}
    \end{minipage}%
      \begin{minipage}{0.1\textwidth}
        \centering
        \includegraphics[width=\textwidth]{ 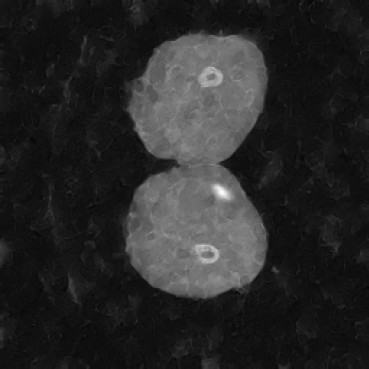}
    \end{minipage}%
      \begin{minipage}{0.1\textwidth}
        \centering
        \includegraphics[width=\textwidth]{ 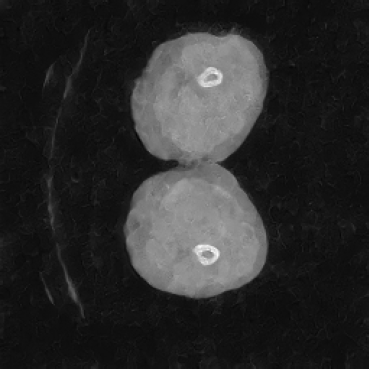}
    \end{minipage}%
      \begin{minipage}{0.1\textwidth}
        \centering
        \includegraphics[width=\textwidth]{ 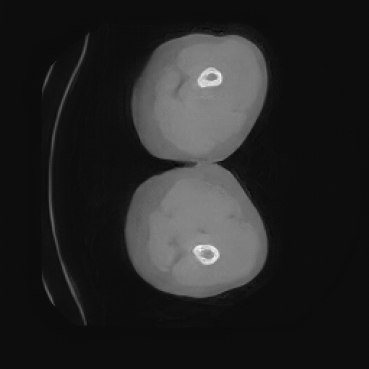}
    \end{minipage}%
    
    \caption{Conditional AutoPET progressive generation}
    \label{fig:progressive}
\end{figure}

\section{More Generated Images Slices}

\begin{figure}[]
    \centering
    \captionsetup{list=no}
       \includegraphics[width=0.2\textwidth, angle=180]{ 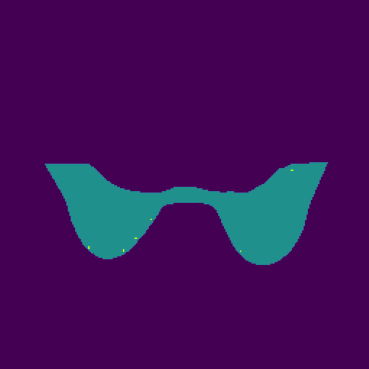} 
      \includegraphics[width=0.2\textwidth, angle=180]{ 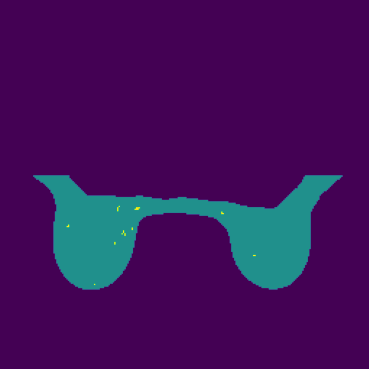} 
    \includegraphics[width=0.2\textwidth, angle=180]{ 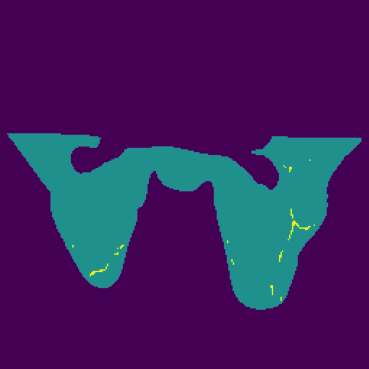} 
      \includegraphics[width=0.2\textwidth, angle=180]{ 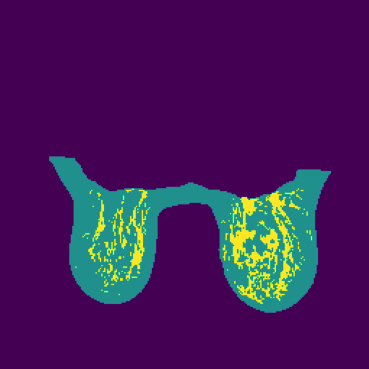} 

    \includegraphics[width=0.2\textwidth, angle=180]{ 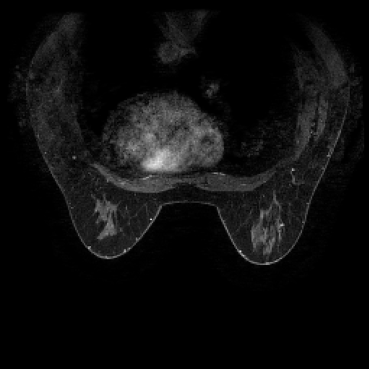} 
      \includegraphics[width=0.2\textwidth, angle=180]{ 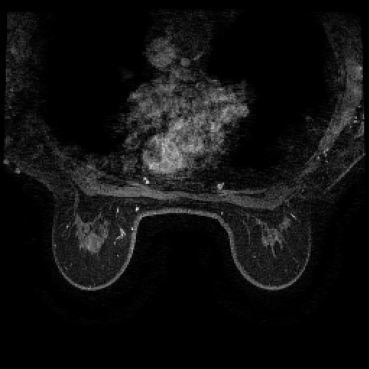} 
    \includegraphics[width=0.2\textwidth, angle=180]{ 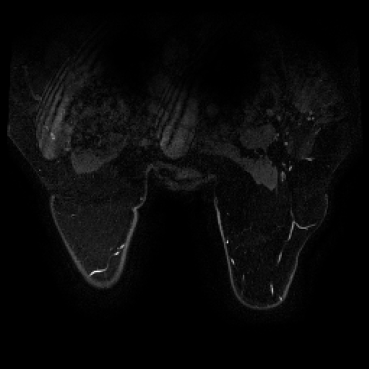} 
      \includegraphics[width=0.2\textwidth, angle=180]{ 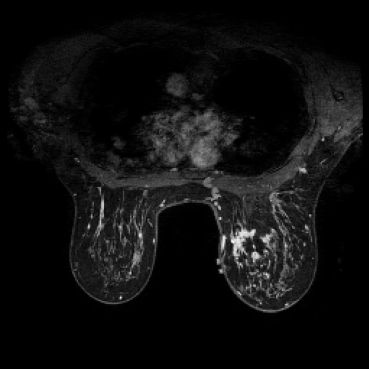} 

      \includegraphics[width=0.2\textwidth, angle=180]{ 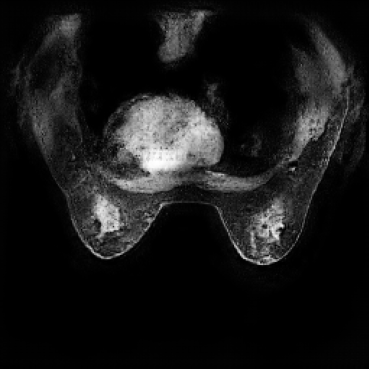} 
      \includegraphics[width=0.2\textwidth, angle=180]{ 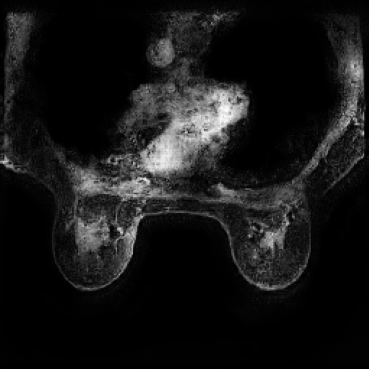} 
    \includegraphics[width=0.2\textwidth, angle=180]{ 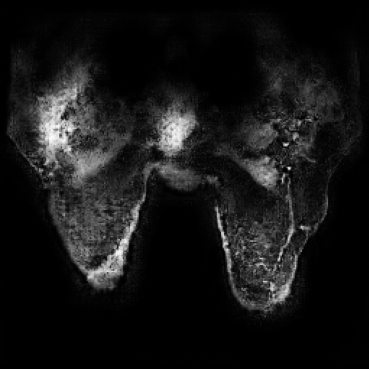} 
      \includegraphics[width=0.2\textwidth, angle=180]{ 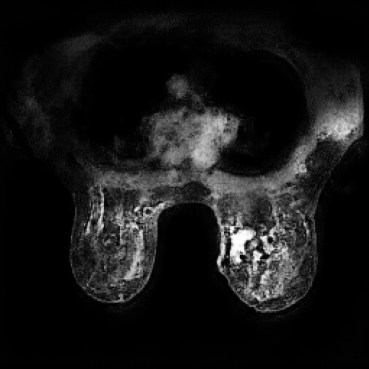} 

        \includegraphics[width=0.2\textwidth, angle=180]{ 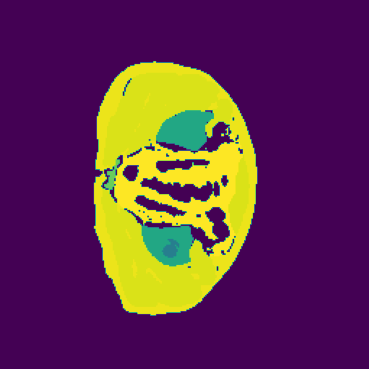} 
      \includegraphics[width=0.2\textwidth, angle=180]{ 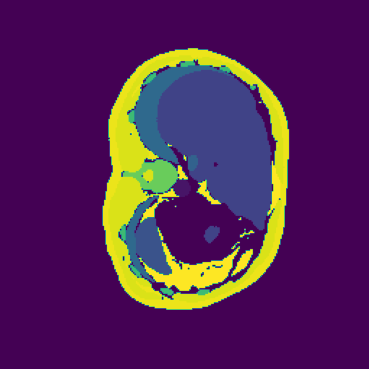} 
    \includegraphics[width=0.2\textwidth, angle=180]{ 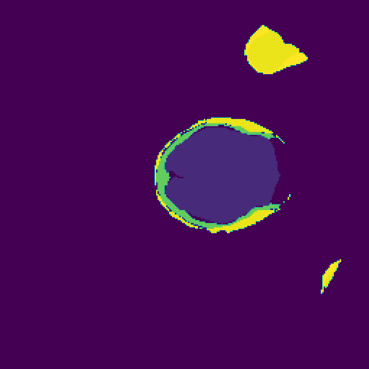} 
      \includegraphics[width=0.2\textwidth, angle=180]{ 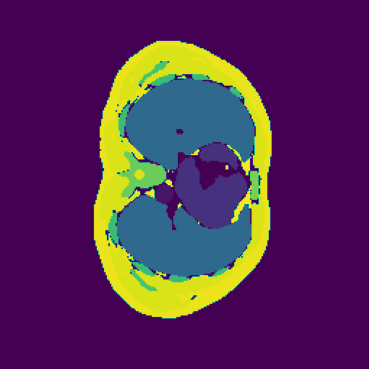} 

    \includegraphics[width=0.2\textwidth, angle=180]{ figures/diffusion_autopet/10_image_slice_16.png} 
      \includegraphics[width=0.2\textwidth, angle=180]{ 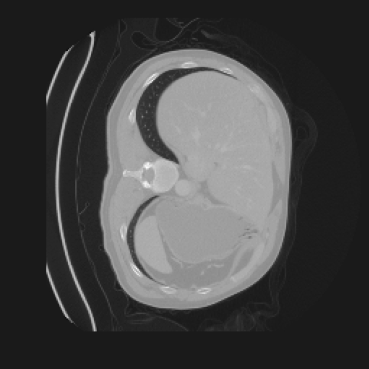} 
    \includegraphics[width=0.2\textwidth, angle=180]{ 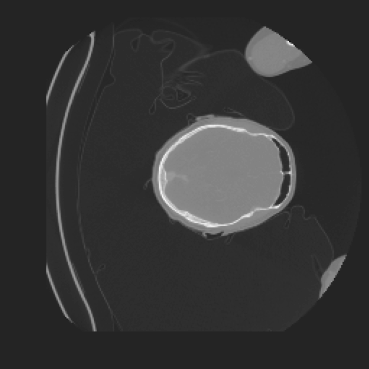} 
      \includegraphics[width=0.2\textwidth, angle=180]{ 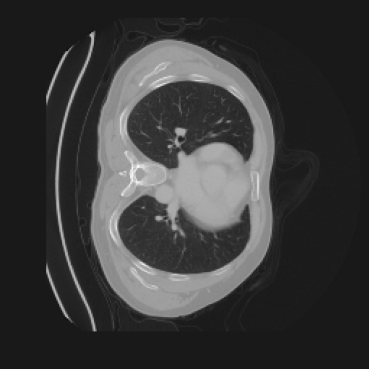} 

      \includegraphics[width=0.2\textwidth, angle=180]{ figures/diffusion_autopet/10_generated_slice_16.png} 
      \includegraphics[width=0.2\textwidth, angle=180]{ 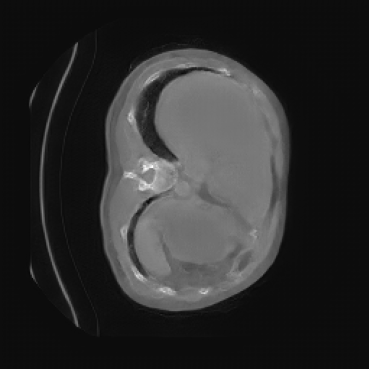} 
    \includegraphics[width=0.2\textwidth, angle=180]{ 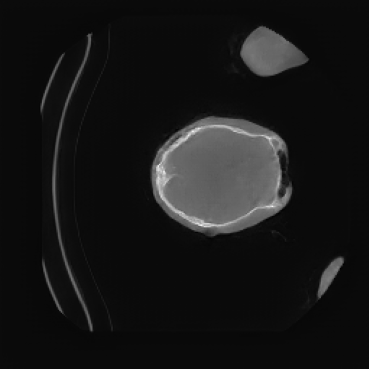} 
      \includegraphics[width=0.2\textwidth, angle=180]{ 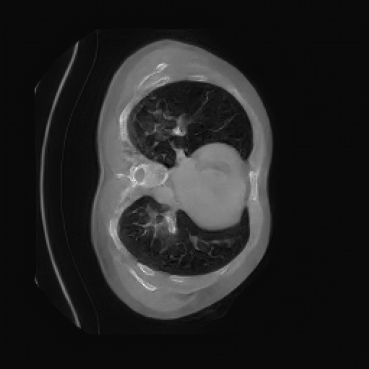}

    \label{fig:duke_part}
\end{figure}

\FloatBarrier 

\begin{figure}[H]
    \ContinuedFloat
    \captionsetup{list=no}
    \centering
  
        \includegraphics[width=0.2\textwidth, angle=180]{ 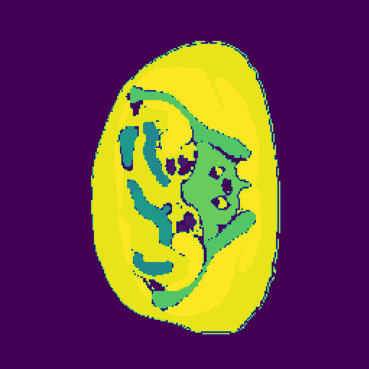} 
      \includegraphics[width=0.2\textwidth, angle=180]{ 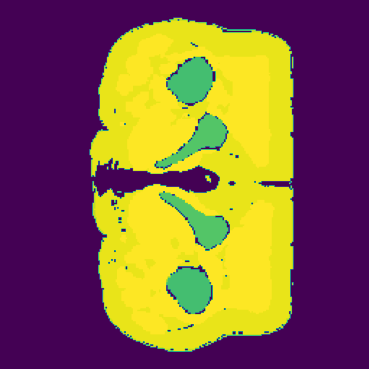} 
    \includegraphics[width=0.2\textwidth, angle=180]{ 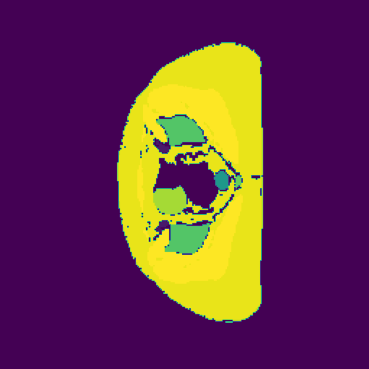} 
      \includegraphics[width=0.2\textwidth, angle=180]{ 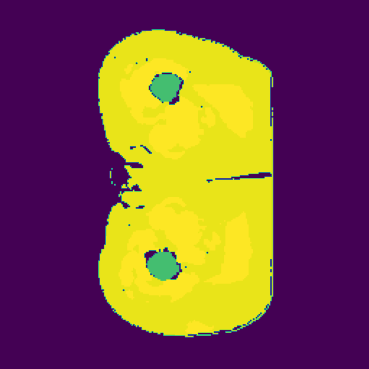} 

    \includegraphics[width=0.2\textwidth, angle=180]{ 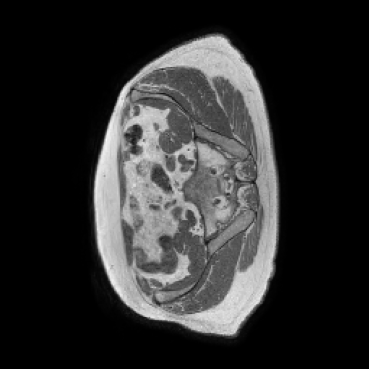} 
      \includegraphics[width=0.2\textwidth, angle=180]{ 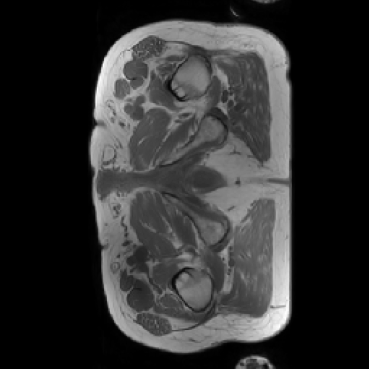} 
    \includegraphics[width=0.2\textwidth, angle=180]{ 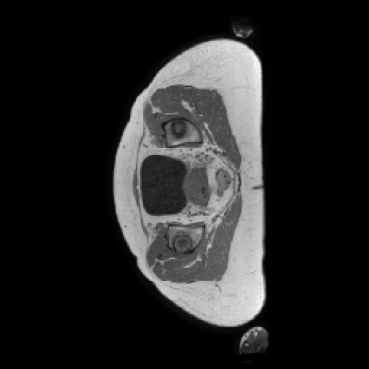} 
      \includegraphics[width=0.2\textwidth, angle=180]{ 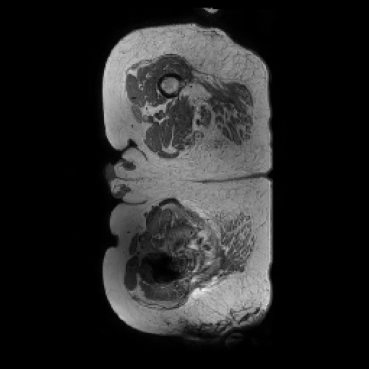} 

      \includegraphics[width=0.2\textwidth, angle=180]{ 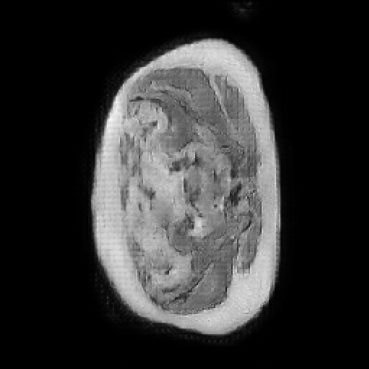} 
      \includegraphics[width=0.2\textwidth, angle=180]{ 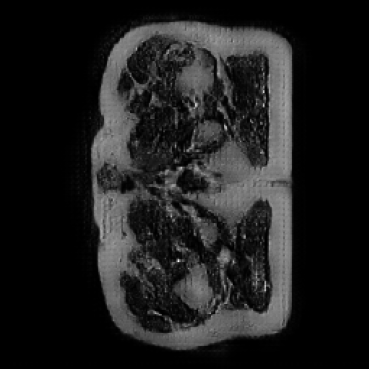} 
    \includegraphics[width=0.2\textwidth, angle=180]{ 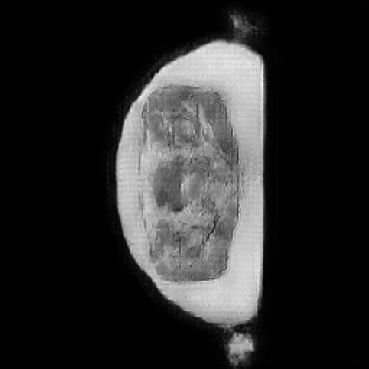} 
      \includegraphics[width=0.2\textwidth, angle=180]{ 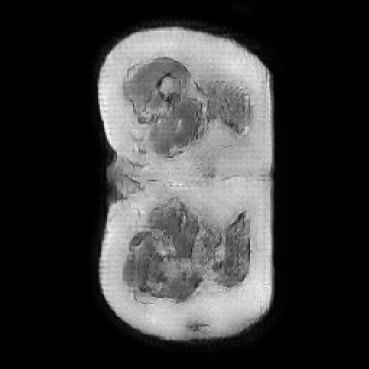} 

    \caption{Duke Breast (top columns in last page), AutoPET (bottoms columns in last page) and SynthRAD2023 (top columns in this page) results, The first, fourth and seventh columns are semantic maps. The second, fifth and eighths columns are real image slices. The third, sixth and ninth columns are synthetic image slices generated by Med-LSDM}
\end{figure}

\end{document}